\documentclass[prb,aps,twocolumn,superscriptaddress]{revtex4-2}
\usepackage{graphicx,xcolor}
\usepackage{amsthm}
\usepackage{amsfonts}
\usepackage{algorithmic}
\usepackage{enumerate}
\usepackage{latexsym}
\usepackage{amsmath}
\usepackage{amssymb}
\usepackage{subfig}
\usepackage[colorlinks=true,citecolor=blue,linkcolor=blue]{hyperref}

\usepackage{dcolumn}
\usepackage{bm}

\usepackage[T1]{fontenc}
\usepackage{mathptmx}

\usepackage{natbib}

\usepackage{caption}

\begin{document}

\title{Tuning superconductivity and charge density wave order by next-nearest-neighbor hopping integral in honeycomb Holstein model}
\author{Hongxing Liu}
\affiliation{School of Physics and Astronomy, Beijing Normal University, and Key Laboratory of Multiscale Spin Physics (Beijing Normal University), Ministry of Education, Beijing 100875, China\\}
\author{Lufeng Zhang}
\affiliation{School of Physical Science and Technology, Beijing University of Posts and Telecommunications, Beijing 100876, China\\}
\author{Tianxing Ma}
\email{txma@bnu.edu.cn}
\affiliation{School of Physics and Astronomy, Beijing Normal University, and Key Laboratory of Multiscale Spin Physics (Beijing Normal University), Ministry of Education, Beijing 100875, China\\}

\begin{abstract}
  By using unbiased determinant quantum Monte Carlo simulations, we investigate the interplay between superconductivity  and charge density wave  order in the Holstein model on a honeycomb lattice with next-nearest-neighbor  hopping \(t^{\prime}\). 
  We find that a finite negative \(t^{\prime}\) enhances \(s\)-wave superconducting  pairing susceptibility near the van Hove fillings in the weak electron-phonon coupling regime, while it suppresses superconductivity and promotes charge density wave order at intermediate electron-phonon coupling strengths. 
  The effect of \(t^{\prime}\) on a charge density wave is filling-dependent: It suppresses the charge density wave at half filling but enhances it near the van Hove singularities. 
  A spectral analysis reveals the opening of a gap at low temperatures, highlighting the competitive relationship between superconducting and charge density wave orders mediated by  electron-phonon coupling and tuned by \(t^{\prime}\).
\end{abstract}

\maketitle

\section{Introduction}
Two-dimensional and layered quantum materials often lie in close proximity to multiple ordering instabilities because modest interactions can strongly amplify low-energy particle-hole and particle-particle fluctuations when the band structure provides an enhanced density of states (DOS) or approximate Fermi-surface nesting.
Such intertwined or competing charge density wave (CDW) order and superconductivity (SC) phenomena have been discussed across doped cuprates, kagome superconductors $A$V$_3$Sb$_5$, and transition-metal dichalcogenides (TMDs) \cite{merrittLowenergyPhononsBi2019,pintschoviusElectronPhononCoupling2005,idetaEffectElectronphononCoupling2013,heRapidChangeSuperconductivity2018,kendzioraUnconventionalSuperconductivityObserved2001,luoElectronicNatureCharge2022,angRealspaceCoexistenceMelted2012,liuSuperconductivityInducedSedoping2013,choUsingControlledDisorder2018}.
Electron-phonon coupling (EPC) offers a controllable route to this competition: The phonon-mediated interaction can promote on-site $s$-wave pairing and SC, while simultaneously enhancing charge correlations and stabilizing CDW order at wave vectors selected by the underlying Fermi-surface geometry \cite{bardeenTheorySuperconductivity1957,zhuClassificationChargeDensity2015,grunerDynamicsChargedensityWaves1988}.
Although mass renormalization and superconductivity can also arise from nonphononic channels, a central issue in EPC-relevant settings is how band-structure tuning controls the balance between SC and CDW tendencies.
In particular, on the honeycomb lattice the next-nearest-neighbor (NNN) hopping $t'$ reshapes its energy dispersion by breaking particle-hole symmetry and distorting the Fermi surface, which shifts the energy associated with van Hove physics and modifies nesting conditions. 
In certain parameter ranges of $t'/t$, it can even generate additional saddle points that further enhance the low-energy DOS \cite{linQuantumMonteCarlo2015}.

In this context, the Holstein model \cite{WOS:000086983000016} provides a minimal yet versatile framework for studying local EPC and the resulting competition between CDW order and SC.
A large body of analytical and unbiased numerical work has established that, on many lattices and especially at or near half filling where Fermi-surface geometry favors strong particle-hole fluctuations, Holstein-type models commonly develop robust long-range CDW order over a broad range of EPC strengths $\lambda_{D}$ \cite{zhangChargeOrderHolstein2019,fengInterplayFlatElectronic2020,nosarzewskiSuperconductivityChargeDensity2021,liEnhancementSuperconductivityFrustrating2019}.
Geometric frustration can substantially reshape this balance. For example, on the kagome lattice CDW order has been reported only in restricted windows of filling and coupling \cite{bradleyChargeOrderKagome2023}, highlighting increased sensitivity to band-structure details.
Beyond varying $\lambda_{D}$, several additional knobs have been explored to tune or disentangle CDW and SC tendencies, including phonon dispersion \cite{costaPhononDispersionCompetition2018}, spatially inhomogeneous EPC \cite{mengSupersolidPhaseDiluted2024}, carrier doping \cite{yingChargeDensityWave2024}, and chemical disorder \cite{xiaoChargeDensityWave2021}.
Among the cleanest and most direct routes is band-structure engineering via NNN hopping $t'$, which modifies the Fermi-surface geometry and low-energy DOS, thereby affecting the relative strength of particle-hole and pairing channels.
On a square lattice, for instance, $t'$ has been shown to enhance $s$-wave pairing away from half filling but suppress CDW tendencies close to half filling \cite{vekicChargedensityWavesSuperconductivity1992}.

Motivated by the successful synthesis of graphene in 2004 \cite{novoselovElectricFieldEffect2004}, the honeycomb lattice has become a paradigmatic platform where band-structure features, such as Dirac cones at half filling and van Hove physics upon doping, can be tuned in a controlled manner.
While the EPC-driven physics of Dirac fermions at half filling and in the lightly doped regime is relatively well understood, including a finite-temperature Ising transition into a CDW phase above a critical coupling \cite{zhangChargeOrderHolstein2019,yingChargeDensityWave2024}, the situation becomes far less settled when the filling is tuned toward the van Hove singularity (vHS).
In this regime, where the DOS is strongly enhanced and $t'$ plays an increasingly important role in reshaping the low-energy electronic structure \cite{maControllabilityFerromagnetismGraphene2010,linQuantumMonteCarlo2015,jiaPairingHubbardModel2022,liMixtureNearestNextnearestneighbor2022,shenTransitionHalffilledStripe2024}, how $t'$ tunes the interplay between SC and CDW remains an open question.

In this work, we carry out hybrid determinant quantum Monte Carlo (DQMC) \cite{10.21468/SciPostPhysCodeb.29, 10.21468/SciPostPhysCodeb.29-r0.3} to obtain high-quality unbiased data.
We use these results to determine how NNN hopping $t'$ tunes the intertwined CDW and $s$-wave pairing tendencies in the doped honeycomb Holstein model.
Furthermore, we apply differential evolution analytic continuation (DEAC) \cite{10.21468/SciPostPhysCodeb.39,10.21468/SciPostPhysCodeb.39-r1.1} to the imaginary-time fermionic Green's function measured in DQMC and extract the single-particle spectral function $A(\mathbf{k},\omega)$.
This may offer a possible point of contact with future spectroscopic measurements such as angle-resolved photoemission spectroscopy (ARPES).
We also note that the same continuation framework can be applied to a bosonic Green's function in future work, which would enable qualitative comparisons to inelastic scattering probes of phonon spectra.

\section{Model and Method}

The Holstein Hamiltonian on the honeycomb lattice is defined as
\begin{equation}
  \begin{aligned}
    \hat{H} = &-t \sum_{\textbf{i}\mathbf{\eta} \sigma}
    \hat{a}^{\dagger}_{\textbf{i}\sigma}\hat{b}^{\phantom{\dagger}}_{\textbf{i}+\mathbf{\eta}\sigma}
    - t^{\prime} \sum_{\textbf{i}\mathbf{\gamma}\sigma} \left( \hat{a}^{\dagger}_{\textbf{i}\sigma}\hat{a}^{\phantom{\dagger}}_{\textbf{i}+\mathbf{\gamma}\sigma} + \hat{b}^{\dagger}_{\textbf{i}\sigma}\hat{b}^{\phantom{\dagger}}_{\textbf{i}+\mathbf{\gamma}\sigma} \right) + \mathrm{h.c.}\\
    &-\mu \sum_{\textbf{i}\sigma} \left( \hat{n}^{\phantom{\dagger}}_{\textbf{i}a \sigma} + \hat{n}^{\phantom{\dagger}}_{\textbf{i}b \sigma} \right)
    +\sum_{\textbf{i} \nu} \left( \frac{\hat{P}^{2}_{\textbf{i} \nu}}{2M} + \frac{1}{2}M\omega_{0}^{2} \hat{X}^{2}_{\textbf{i} \nu} \right)\\
    &+g\sum_{\textbf{i}\sigma} \hat{X}_{\textbf{i} a} \left( \hat{n}^{\phantom{\dagger}}_{\textbf{i}a\sigma} - \frac{1}{2} \right) +
    \hat{X}_{\textbf{i} b} \left( \hat{n}^{\phantom{\dagger}}_{\textbf{i}b\sigma} - \frac{1}{2} \right)
  \end{aligned}
  \label{eq:Hol_Ham}
\end{equation}
which describes conduction electrons with spin $\sigma$ ($\sigma = \uparrow, \downarrow$) locally coupled to the dispersionless phonon mode on each site through an electron-phonon interaction $g$ \cite{WOS:000086983000016}, and hopping to the nearest-neighbor (NN) and NNN sites connected by vectors $\eta$ and $\gamma$.
Here, $\hat{a}^{\phantom{\dagger}}_{\textbf{i}\sigma}$ ($\hat{a}^{\dagger}_{\textbf{i}\sigma}$) annihilates (creates) electrons at site $\textbf{i}$ with spin $\sigma$ on sublattice A, $\hat{b}^{\phantom{\dagger}}_{\textbf{i}\sigma}$ ($\hat{b}^{\dagger}_{\textbf{i}\sigma}$) annihilates (creates) electrons at site $\textbf{i}$ with spin $\sigma$ on sublattice B, while $\hat{n}^{\phantom{\dagger}}_{\textbf{i}a\sigma}=\hat{a}^{\dagger}_{\textbf{i}\sigma}\hat{a}^{\phantom{\dagger}}_{\textbf{i}\sigma}$ and $\hat{n}^{\phantom{\dagger}}_{\textbf{i}b\sigma}=\hat{b}^{\dagger}_{\textbf{i}\sigma}\hat{b}^{\phantom{\dagger}}_{\textbf{i}\sigma}$ are the corresponding number operators.
The parameters $t$ and $t^{\prime}$ are the NN and NNN hopping energy, respectively, and $\mu$ is the chemical potential.
On each unit cell $\textbf{i}$, the local oscillators of fixed frequency $\omega_{0}$ are set with a normalized mass $M=1$ and corresponding phonon position and momentum operators with sublattice index $\nu$ ($\nu=a,b$), $\hat{X}^{\phantom{2}}_{\textbf{i}\nu}$ and $\hat{P}^{\phantom{2}}_{\textbf{i}\nu}$, respectively.
The strength of electrons coupled to the local phonon could be evaluated by a dimensionless parameter $\lambda_{D} = 2g^{2} / \omega_{0}^{2}W$, where $W$ is the bandwidth of a honeycomb lattice with finite $t^{\prime}$. In this study, we set $t=1$ as the energy unit. 
Meanwhile, in order to capture the competition relationship between SC and CDW order, we set $\omega_{0}=|t|$ to involve the retardation of the phonon-mediated interaction. 
This value of $\omega_{0}$ is neither adiabatic nor antiadiabatic and the DQMC treatment fully includes vertex corrections, thus making it accessible to isolate how $t^{\prime}$ reshapes the competing CDW and pairing tendencies from the perspective of theoretical exploration.

This model is sign-problem-free because it is spin independent. 
The partition function is given by
\begin{equation}
  Z = \int dx \; e^{-E_{lat}\Delta \tau} \; \prod_{\sigma=\uparrow,\downarrow} \det M^{\phantom{\dagger}}_{\sigma}
  \label{eq:Z}
\end{equation}
where $E_{\textrm{lat}}$ is determined by the phonon fields and $ M^{\phantom{\dagger}}_{\sigma}$ is the spin-$\sigma$ fermionic matrix \cite{johnstonDeterminantQuantumMonte2013}. 
For the real NN and NNN hopping integrals we choose, the fermion matrices for spin-up and spin-down electrons are identical regardless of any phonon fields, which makes the Monte Carlo weight always positive.

As for the tight-binding model of a honeycomb lattice considering the NN and NNN hopping terms, the energy bands of Eq.~\ref{eq:Hol_Ham} are given by
\begin{equation}
  \begin{aligned}
    E_{\pm} &= \pm t \sqrt{3+f(\textbf{k})} - t^{\prime}f(\textbf{k}),\\
    f(\textbf{k}) &= 2\cos\left( \sqrt{3} k_{y} a \right) + 4\cos\left( \frac{\sqrt{3}}{2} k_{y} a \right) \cos\left( \frac{3}{2}k_{x} a \right),
  \end{aligned}
  \label{eq:E(k)}
\end{equation}
where the plus and minus signs correspond to the upper and lower band, respectively.
When a finite $t^{\prime}$ is introduced, the particle-hole symmetry is broken, thus making the DOS asymmetric about the Fermi level \cite{linQuantumMonteCarlo2015}. For $|t^{\prime}|<|t|/6$, two vHSs are located at $\langle n \rangle = 3/4$ and $5/4$, corresponding to $E = \pm t - 2t^{\prime}$, while a third vHS near the former two peaks appears for $|t^{\prime}|\ge|t|/6$ \cite{maControllabilityFerromagnetismGraphene2010}. As doping becomes more heavy, the effect of $t^{\prime}$ should be taken into consideration and its value ranges from 0.02 to 0.2 according to an \textit{ab initio} calculation \cite{liGWStudyMetalinsulator2002}.

In the Holstein model, the effective electron-electron interaction mediated by a phonon, $V^{\textrm{eff}}_{e-e} (\omega) = -2\lambda^{2}\omega_{0}/(\omega_{0}^{2} - \omega^{2})$ is determined in second-order perturbation theory \cite{bergerTwodimensionalHubbardHolsteinModel1995}, where $\lambda = g\sqrt{2\omega_{0}}$ and the minus sign implies that this interaction is attractive. 
With these vHS peaks tuned by $t^{\prime}$ as well as a local phonon-mediated interaction taken into consideration, the properties of the honeycomb lattice are manifested carefully by the following quantities.  To study the SC and CDW phases, we define and measure the $s$-wave channel superconducting pairing susceptibility $P_{s}$ and the CDW structure factor $S_{\text{CDW}}$.
The SC phase is depicted by
\begin{equation}
  P_{s} = \frac{1}{N}\int_{0}^{\beta} d\tau \langle \hat{\Delta}(\tau)\hat{\Delta}^{\dagger}(0) \rangle,
  \label{eq:Ps}
\end{equation}
where $\hat{\Delta}(\tau) = \sum^{\phantom{\dagger}}_{\textbf{i}} \hat{a}^{\phantom{\dagger}}_{\textbf{i} \downarrow}(\tau)\hat{a}^{\phantom{\dagger}}_{\textbf{i} \uparrow}(\tau) + \hat{b}^{\phantom{\dagger}}_{\textbf{i} \downarrow}(\tau)\hat{b}^{\phantom{\dagger}}_{\textbf{i} \uparrow}(\tau)$,
$\hat{a}^{\phantom{\dagger}}_{\textbf{i} \sigma}(\tau) = e^{\hat{H}\tau}\hat{a}^{\phantom{\dagger}}_{\textbf{i}\sigma} (0) e^{-\hat{H}\tau}$,
$\hat{b}^{\phantom{\dagger}}_{\textbf{i} \sigma}(\tau) = e^{\hat{H}\tau}\hat{b}^{\phantom{\dagger}}_{\textbf{i}\sigma} (0) e^{-\hat{H}\tau}$, and $\tau\in[0,\beta)$.
Here $N=2\times L^{2}$ is the total number of lattice sites. 
To provide a more sensitive measurement on the SC phase, we use an integrated correlation function along imaginary time slices for the pairing order, and to alleviate the systematic Trotter errors from the sampling, we set the discretization mesh $\Delta\tau = 0.05$.

We describe the CDW phase by a charge correlation function in real space
\begin{equation}
  \begin{aligned}
    c^{\phantom{\dagger}}_{\nu \nu^{\prime}}(\textbf{r}) = \langle ( \hat{n}^{\phantom{\dagger}}_{\textbf{i} \nu \uparrow} + \hat{n}^{\phantom{\dagger}}_{\textbf{i} \nu \downarrow} ) ( \hat{n}^{\phantom{\dagger}}_{\textbf{i}+\textbf{r} \nu^{\prime}\uparrow} + \hat{n}^{\phantom{\dagger}}_{\textbf{i}+\textbf{r} \nu^{\prime}\downarrow} ) \rangle
  \end{aligned}
  \label{eq:CDW}
\end{equation}
and its Fourier transform, the CDW structure factor,
\begin{equation}
  \begin{aligned}
    S_{\text{CDW}}(\textbf{k}) = \sum^{\phantom{\dagger}}_{\textbf{r}} (-1)^{\textbf{r}}c(\textbf{r}).
  \end{aligned}
  \label{eq:SCDW}
\end{equation}
The $-1$ phase accesses the staggered pattern of the charge ordering between sublattice A and B on the honeycomb lattice. 
Here, it is a convenient way to characterize the CDW phase by calculating this quantity.
Furthermore, we extract the single-particle electron spectral function $A(\textbf{k},\omega)$ from the unbiased DQMC data. Combined with the single-particle electron spectral functions
\begin{equation}
  G(\textbf{k},\tau) = \langle \hat{c}(\textbf{k},\tau)\hat{c}^{\dagger}(\textbf{k},0) \rangle = \int_{-\infty}^{\infty} d\omega A(\textbf{k},\omega)\frac{e^{-\tau\omega}}{1+e^{-\beta\omega}},
  \label{eq:Akomega}
\end{equation}
where $G(\textbf{k},\tau)$ is the time-dependent Green's function in $\textbf{k}$ space, $\beta = l\Delta\tau$ is the inverse temperature and $l$ is the number of imaginary time slices corresponding to different finite temperatures. 
If $A(\textbf{k}, \omega)$ decreases to zero with temperature being lowered, it indicates the formation of a CDW gap, which roughly suggests the transition temperature $T_{c}$. 
Moreover, the spectral functions at different wave vectors $\textbf{k}$ could give a more detailed description of the CDW phase, especially at some high-symmetry points.

\section{Results and Discussions}
\begin{figure}[t]
  \includegraphics[scale=0.24]{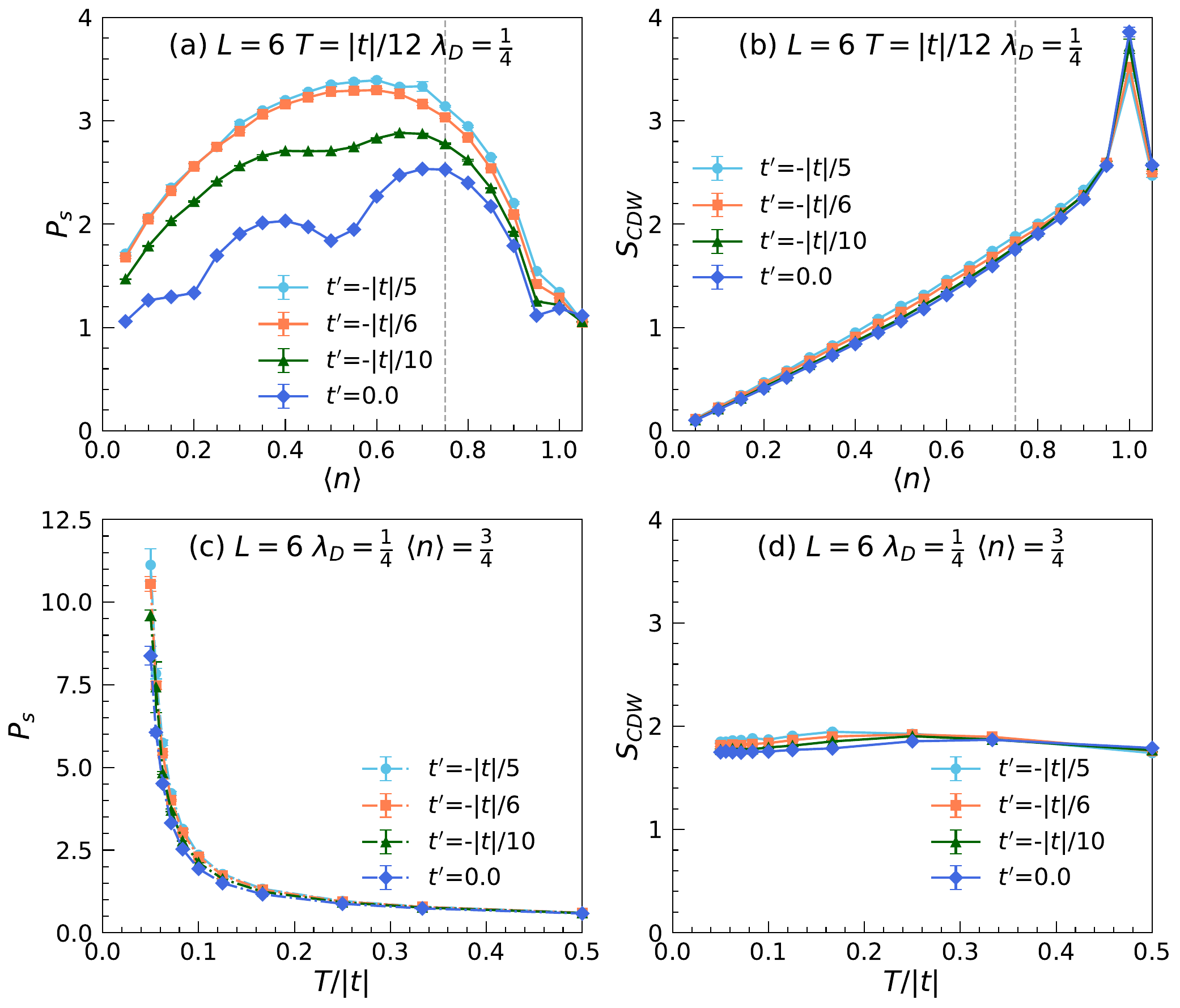}
  \captionsetup{justification=raggedright, singlelinecheck=false}
  \caption{ Pairing susceptibility $P_{s}$ as a function of (a) density $\langle n \rangle$ and (c) temperature $T/|t|$ for different $t^{\prime}$ at $L = 6$ and $\lambda_{D} = 1/4$. The CDW structure $S_{\text{CDW}}$ as (b) a function of density $\langle n \rangle$ and (d) temperature $T/|t|$ for different $t^{\prime}$ at $L = 6$ and $\lambda_{D} = 1/4$. The gray dashed lines in (a) and (b) mark the vHS filling at $\langle n \rangle = 3/4$. }
  \label{FIG1}
\end{figure}

First, we measured the $s$-wave SC pairing susceptibility \( P_s \) and CDW structure factor \( S_{\text{CDW}} \) as functions of electron density \( \langle n \rangle \) and temperature \( T/|t| \), considering different strengths of the NNN hopping integral \( t' \), with the lattice size fixed at \( L = 6 \) and the electron-phonon coupling  strength set to \( \lambda_D = 1/4 \). As shown in Fig.~\ref{FIG1}(a), \( t' \) enhances \( P_s \) over a broad range of fillings where \( \langle n \rangle < 1 \), compared to the case without NNN hopping (\( t' = 0 \)). This indicates that for certain fillings, a much higher SC pairing susceptibility can be achieved at lower temperatures—a trend further verified in Fig.~\ref{FIG1}(c).  Additionally, when \( t' = 0 \), a more pronounced drop in \( P_s \) is observed around \( \langle n \rangle = 1/2 \) compared to other \( t' \) values; this behavior is attributed to finite-size effects.

We further calculated \( P_s \) across a wide temperature range, from \( T = |t|/5 \) to \( T = |t|/20 \) at \( \langle n \rangle = 3/4 \) , a filling corresponding to the van Hove singularity of the honeycomb lattice. 
When the temperature is below \( T = |t|/4 \), the effect of \( t' \) in enhancing \( P_s \) becomes increasingly significant. On the other hand, as shown in Figs.~\ref{FIG1}(b) and \ref{FIG1}(d), \( S_{\text{CDW}} \) remains low in magnitude. This is likely because the SC order takes precedence over the CDW order at the relatively weak EPC strength of \( \lambda_D = 1/4 \).

\begin{figure}[t]
  \includegraphics[scale=0.24]{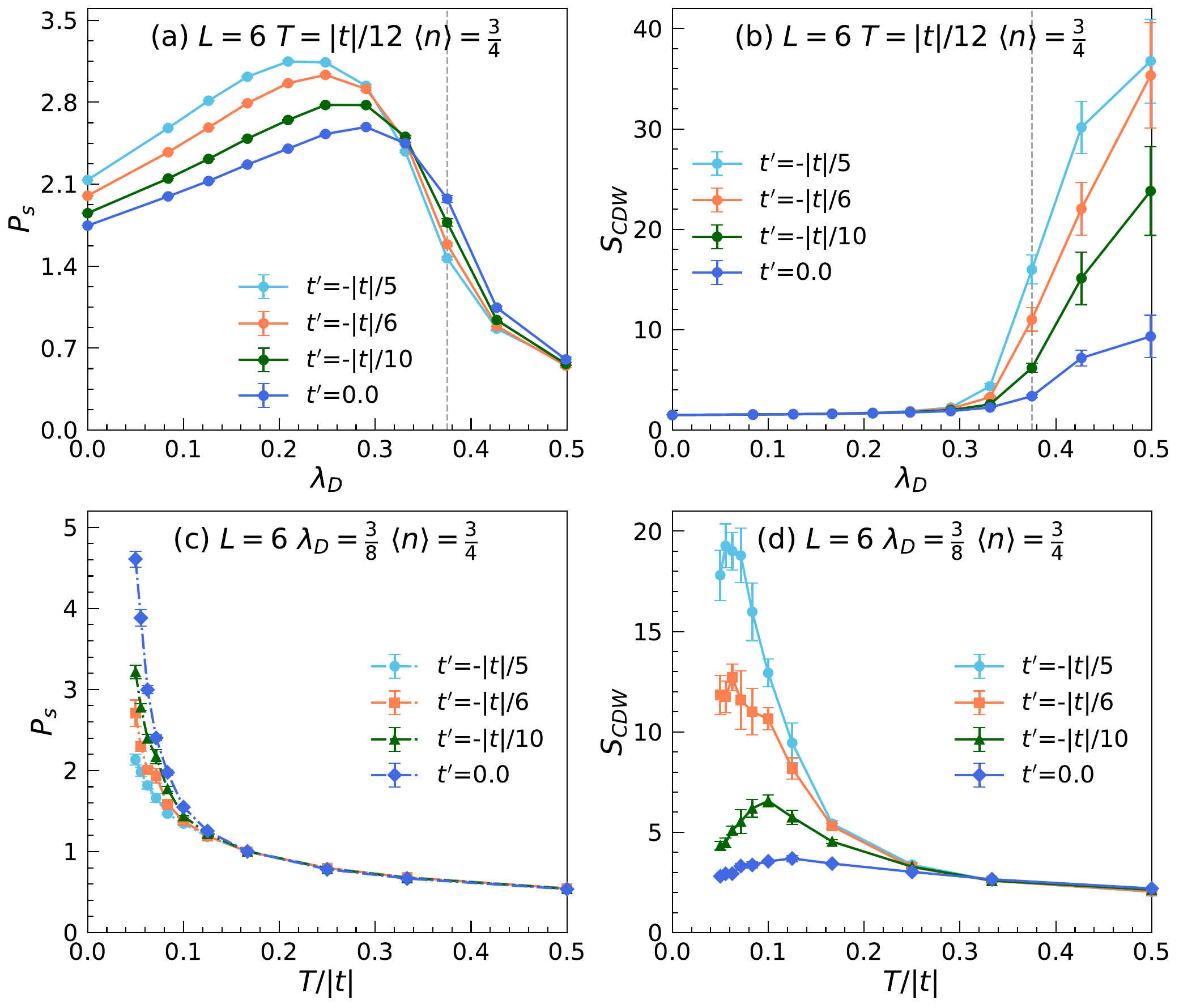}
  \captionsetup{justification=raggedright, singlelinecheck=false}
  \caption{Pairing susceptibility $P_{s}$ as a function of (a) $\lambda_{D}$ and (c) temperature $T/|t|$ for different $t^{\prime}$ at $L = 6, \langle n \rangle = 3/4$. 
  The CDW structure $S_{\text{CDW}}$ as a function of (b) $\lambda_{D}$ and (d) temperature $T/|t|$ for different $t^{\prime}$ at $L = 6, \langle n \rangle = 3/4$.
  The gray dashed lines in (a) and (b) mark EPC strength $\lambda_{D} = 3/8$. }
  \label{FIG2}
\end{figure}

The behavior of the system in the intermediate  EPC regime, $1/3 < \lambda_D < 1/2$, is more complex. Within this regime, $t'$ reduces the $s$-wave SC pairing susceptibility $P_s$ while increasing the CDW structure factor $S_{\text{CDW}}$—a role that is exactly opposite to that of $t'$ in the weak EPC regime, namely $\lambda_D < 1/3$.

As shown in Figs.~\ref{FIG2}(a) and \ref{FIG2}(b), around the gray dashed line, corresponding to the EPC strength $g > t$, $t'$ simultaneously decreases the magnitude of $P_s$ and increases that of $S_{\text{CDW}}$. 
Beyond this critical point, the NNN hopping integral $t'$ weakens the tendency of electron pairs to condense into a superfluid state. 
As the EPC strength increases further, $S_{\text{CDW}}$ becomes dominant while $P_s$ remains low—an effect exemplified by $\lambda_D = 1/2$ [see Figs.~\ref{FIG2}(a) and (b)]. 
At this stage, mediated by local attractive phonon modes, electron pairs tend to localize at specific lattice sites while leaving others empty, thereby driving the system into the CDW phase.

An interesting open question is whether a supersolid phase exists near the vHS within this intermediate EPC regime. Prior to addressing this question, we first revisit the charge order in the half-filled honeycomb lattice and analyze the effect of $t'$ in detail.

\begin{figure}[t]
  \includegraphics[scale=0.32]{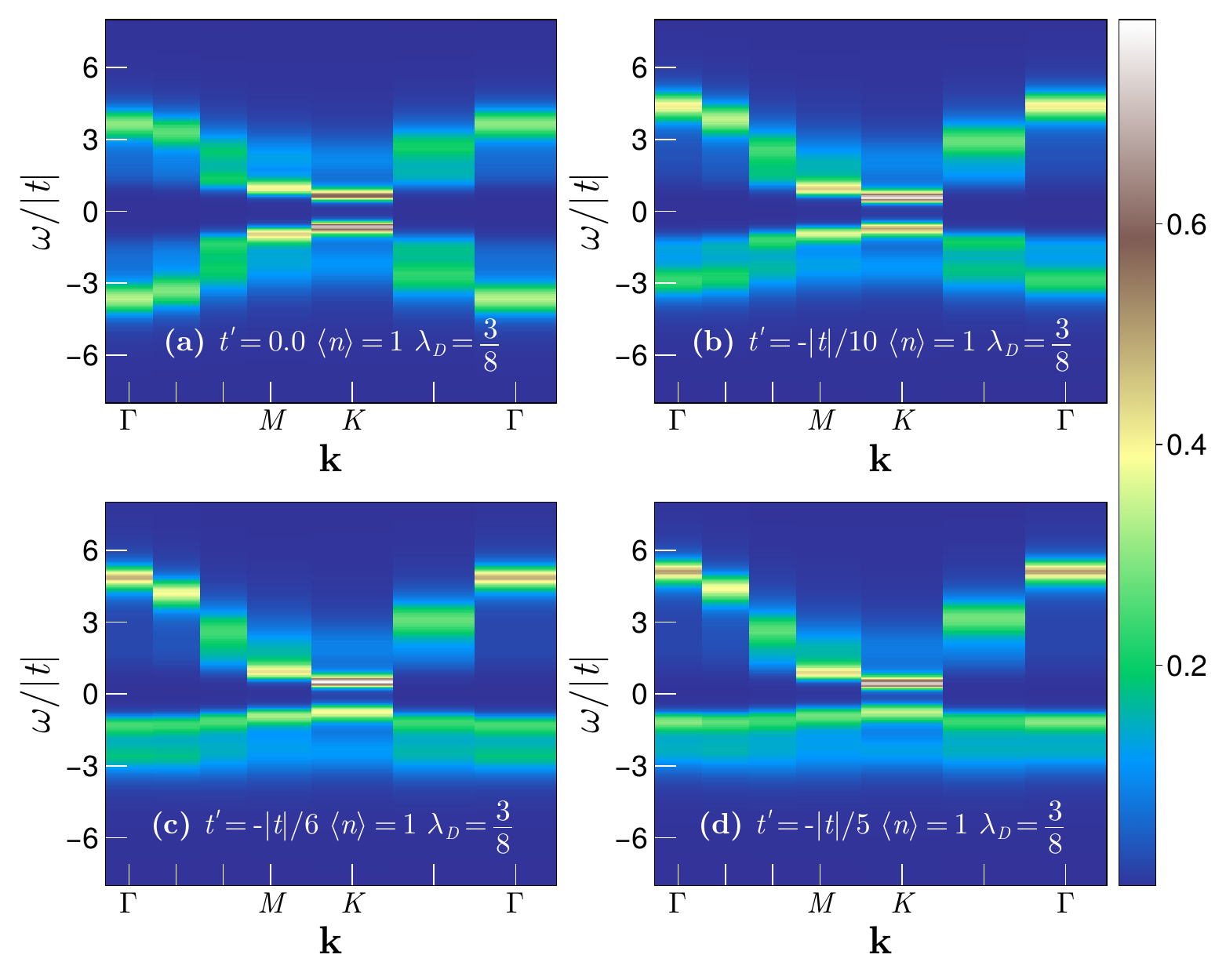}
  \captionsetup{justification=raggedright, singlelinecheck=false}
  \caption{Spectral function $A(\textbf{k},\omega)$ as a function of  electron wave vector $\textbf{k}$ and real frequency $\omega$ at $\langle n \rangle=1$, $\lambda_{D} = 3/8$, $T=|t|/8$, where $t^{\prime} = 0.0$, $-|t|/10$, $-|t|/6$, $-|t|/5$ corresponding to (a)--(d) respectively, characterized by a heat map. }
  \label{FIG3}
\end{figure}

\begin{figure}[t]
  \includegraphics[scale=0.32]{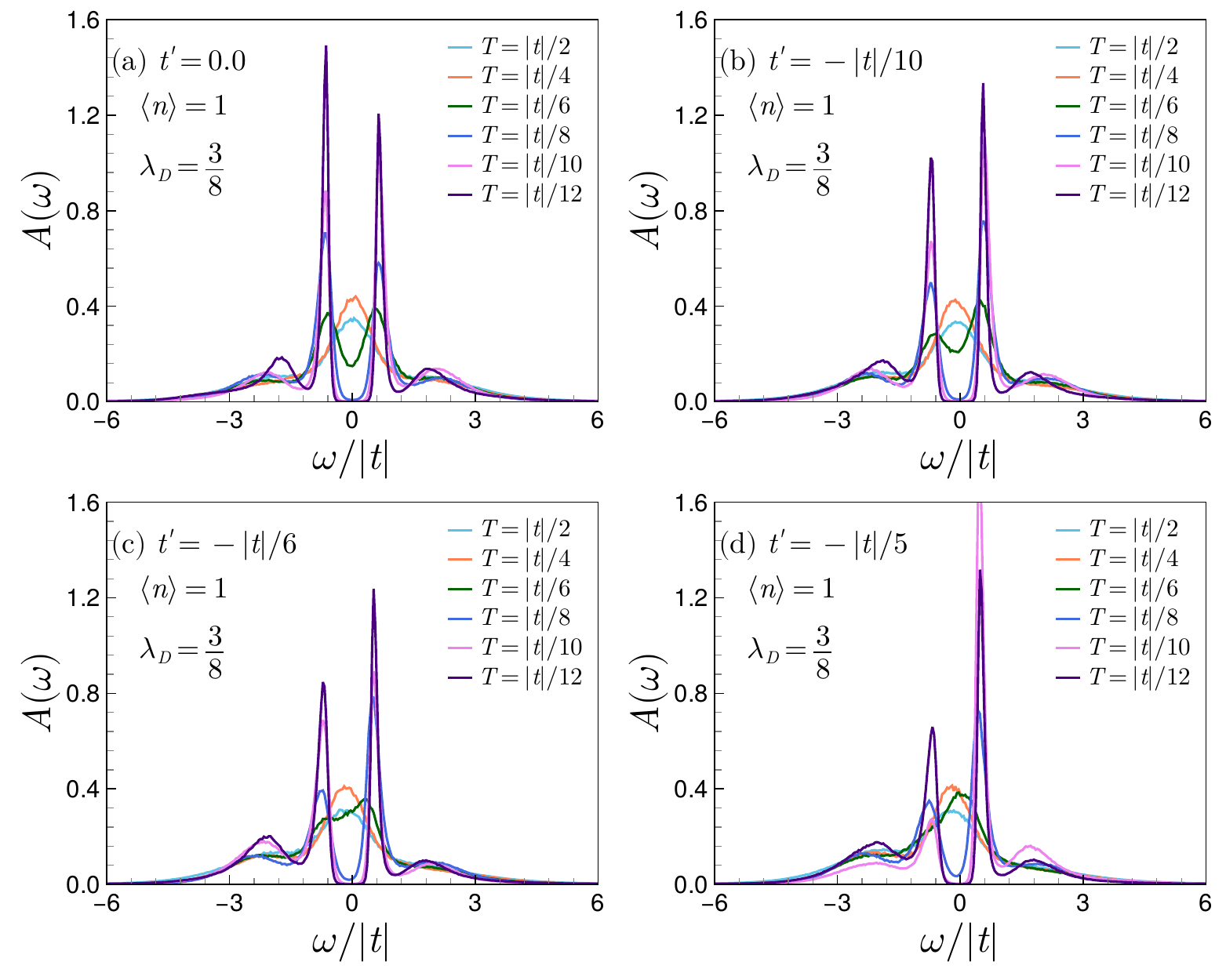}
  \captionsetup{justification=raggedright, singlelinecheck=false}
  \caption{Spectral function $A(\omega)$ as a function of temperature $T$ in a honeycomb lattice, at the Dirac point $K$, fixed $\langle n \rangle = 1$, $ \lambda_{D}=3/8$ where $t^{\prime} = 0.0$, $-|t|/10$, $-|t|/6$, $-|t|/5$, corresponding to (a)--(d), respectively. }
  \label{FIG4}
\end{figure}

The CDW response of the half-filled honeycomb lattice under different EPC strengths has been well studied \cite{zhangChargeOrderHolstein2019}. 
Given the vanishing DOS at $\langle n \rangle = 1$, the CDW phase dominates across a broad range of EPC strengths, starting from $\lambda_D \approx 0.28$. 
To revisit this behavior, we extract the heatmap of the spectral function $A(\textbf{k},\omega)$ at $\lambda_D = 3/8$ and $\langle n \rangle = 1$ along the $\bf k$-space path $\Gamma$-$M$-$K$-$\Gamma$ (see Fig.~\ref{FIG3}), as well as the frequency-dependent spectral function $A(\omega)$ at the Dirac point $K$ (see Fig.~\ref{FIG4}). 
These analyses aim to characterize the robust CDW phase in the absence of the NNN hopping term ($t'=0.0$) and investigate how $t'$ influences different electron wave vectors.

Figure.~\ref{FIG4}(a) ($t'=0.0$) exhibits a clear gap near the Fermi level ($\omega = 0$). 
This gap feature becomes less distinct in Fig.~\ref{FIG4}(b) and vanishes entirely in Figs.~\ref{FIG4}(c) and \ref{FIG4}(d) as $|t'|$ increases. 
This trend can be attributed to the reduced stability of the CDW phase at this relatively small $\lambda_D$. These results demonstrate that $t'$ suppresses the CDW phase in the half-filled honeycomb lattice.


\begin{figure}[t]
  \includegraphics[scale=0.32]{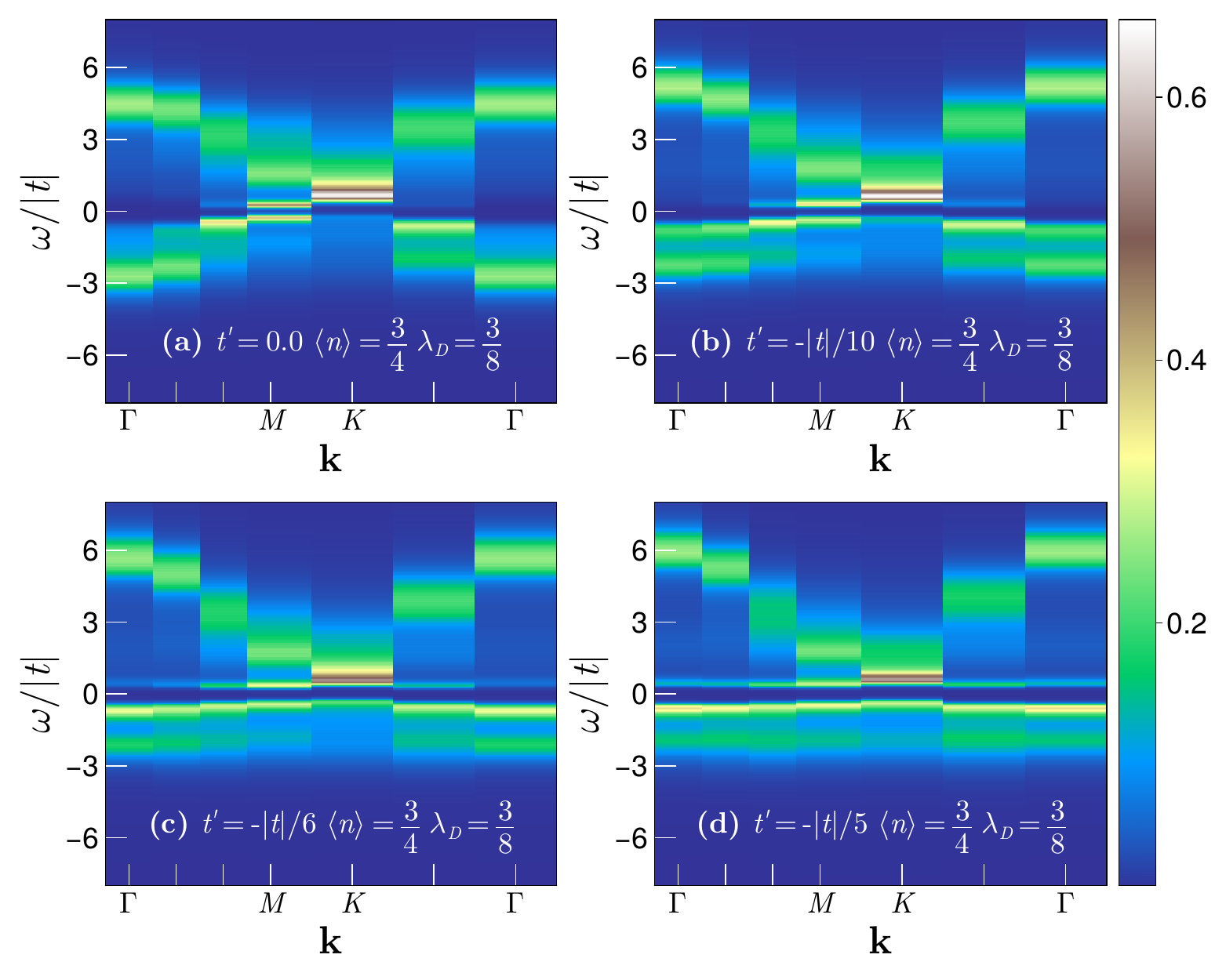}
  \captionsetup{justification=raggedright, singlelinecheck=false}
  \caption{Spectral function $A(\textbf{k},\omega)$ as a function of electron wave vector $\textbf{k}$ and real frequency $\omega$ at fixed $\langle n \rangle = 3/4$, $\lambda_{D} = 3/8$, $T=|t|/20$, where $t^{\prime} = 0.0$, $-|t|/10$, $-|t|/6$, $-|t|/5$ corresponding to (a)--(d) respectively, characterized by a heatmap. }
  \label{FIG5}
\end{figure}

\begin{figure}[t]
  \includegraphics[scale=0.32]{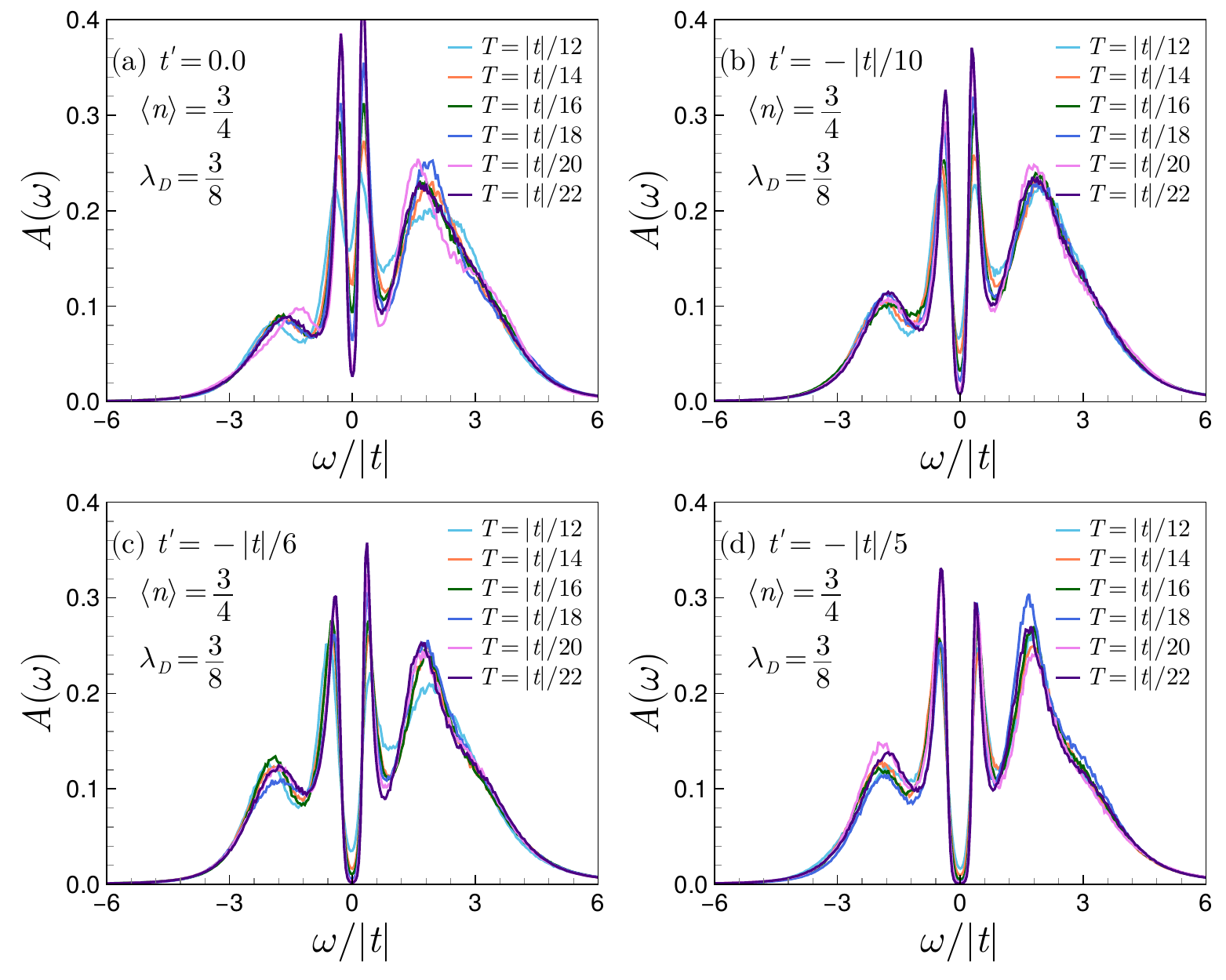}
  \captionsetup{justification=raggedright, singlelinecheck=false}
  \caption{Spectral function $A(\omega)$ as a function of temperature $T$ in a honeycomb lattice at the saddle point $M$, fixed $\langle n \rangle = 3/4$, $\lambda_{D}=3/8$ where $t^{\prime} = 0.0$, $-|t|/10$, $-|t|/6$, $-|t|/5$ corresponding to (a)--(d), respectively. }
  \label{FIG6}
\end{figure}

Turning back to the intermediate EPC regime near the van Hove singularity  of the honeycomb lattice, however, $t'$ plays a crucial role in inducing the CDW phase. 
As shown in Figs.~\ref{FIG5} and \ref{FIG6}, at $\lambda_D = 3/8$ and $\langle n \rangle = 3/4$, a gap gradually emerges near the Fermi level ($\omega = 0$) as $|t'|$ increases.

Specifically, we can confirm this trend in greater detail by focusing on the spectral function at $\textbf{k} = M$ (see Fig.~\ref{FIG6}), where a gap gradually forms as both $|t'|$ and the inverse temperature $|t|/T$ increase. 
By comparing Fig.~\ref{FIG3} with Fig.~\ref{FIG5}, we observe completely opposite effects of $t'$ under different electron fillings: 
In the former case (half filling, $\langle n \rangle = 1$), $t'$ inhibits CDW phase formation, whereas in the latter case (vHS filling, $\langle n \rangle = 3/4$), it promotes the CDW phase.

\begin{figure}[t]
  \includegraphics[scale=0.24]{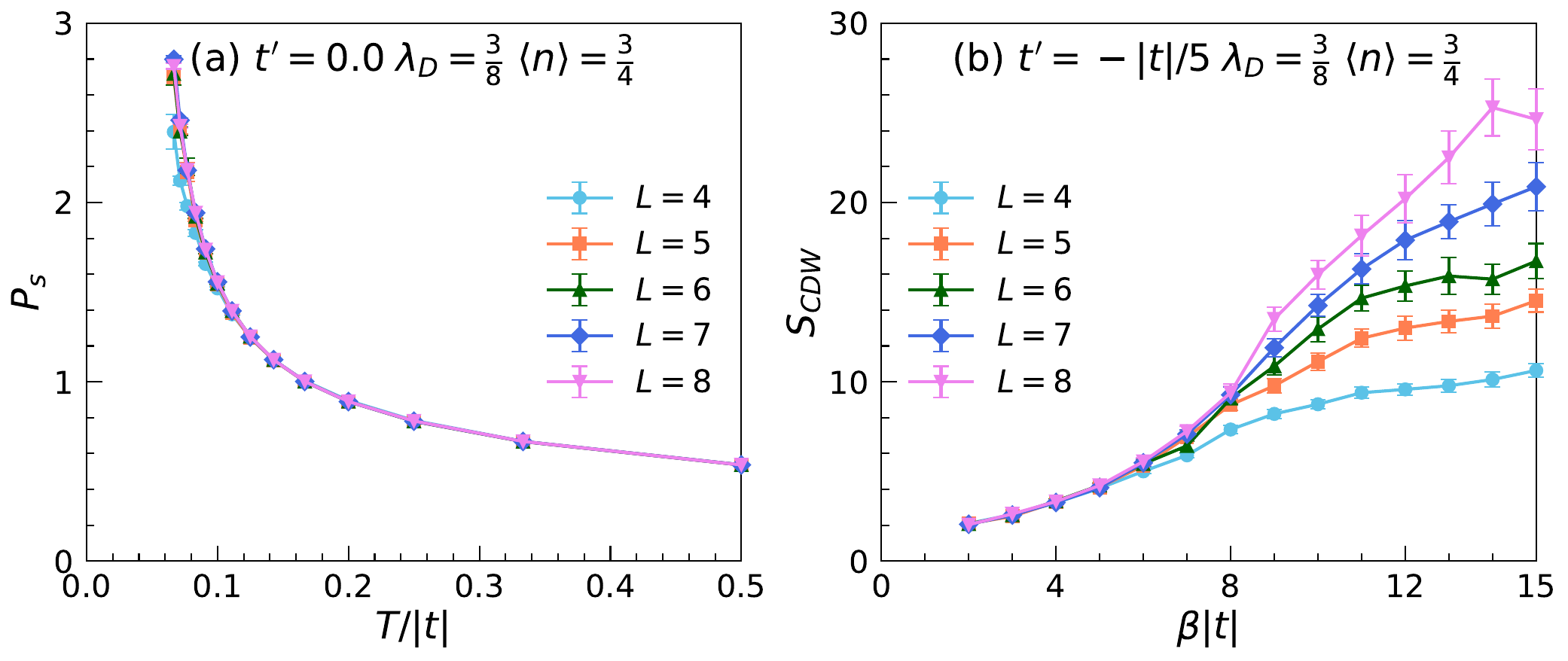}
  \captionsetup{justification=raggedright, singlelinecheck=false}
  \caption{
    Pairing susceptibility $P_{s}$ as a function of (a) temperature $T/|t|$ and (b) CDW structure factor $S_{\text{CDW}}$ as a function of inverse temperature $\beta |t|$ for different lattice sizes $L$ at $\lambda_{D} = 3/8$ and $\langle n \rangle = 3/4$, with $t' = 0.0, -|t|/5$, respectively.
    }
  \label{FIG7}
\end{figure}

Our calculations are mostly performed on the lattice size of $L=6$, and finite-size effects are expected to be particularly important near vHS fillings.
In the thermodynamic limit, the vHS corresponds to a logarithmic divergence of the DOS associated saddle points at $M$, while on finite clusters this divergence is necessarily smeared into a set of discrete peaks.
For the even system sizes considered here the relevant momenta are contained in the discrete Brillouin-zone grid, so the enhanced low-energy weight associated with vHS physics is still captured at the qualitative level.

It is necessary to assess the impact of finite size.
We performed additional simulations on $L=4$--$8$ at representative parameters and examined the size dependence of $P_s$ and $S_{\text{CDW}}$, as shown in Fig.~\ref{FIG7}.
For $t'=0.0$, $P_s$ increases with $L$ for small sizes, but becomes nearly size independent for the largest lattices ($L=7$--$8$) for most values of $\beta |t|$, indicating that the pairing correlation length is smaller than the system size in the accessible temperature range.
For $t' = -|t|/5$, $S_{\text{CDW}}$ is enhanced with increasingly larger lattice size.
While the absolute magnitudes and the detailed temperature scales can shift with $L$, the main qualitative conclusion is robust across the accessible sizes.
Introducing finite $t'$ in the van Hove regime enhances CDW correlations in the intermediate-coupling region and suppresses pairing tendencies relative to $t'=0.0$.
We emphasize, however, that the present system sizes and temperatures do not allow a controlled extraction of critical exponents or precise transition temperatures.
Therefore, throughout this work we focus on robust trends and the relative dominance of CDW versus pairing correlations.

\section{Conclusions}
In summary, we have systematically investigated the modulation of superconducting pairing and charge density wave order by the next-nearest-neighbor hopping term $t^{\prime}$ in the Holstein model on a honeycomb lattice, combining hybrid determinant quantum Monte Carlo simulations and differential evolution analytical continuation for a spectral function analysis.

We find that a finite negative $t^{\prime}$modifies the electronic structure by altering the DOS peaks below half filling, leading to DOS splitting and additional peak formation. 
This structural change enhances the $s$-wave SC pairing susceptibility \( P_s \) near the vHS fillings, particularly in the weak EPC regime (\( \lambda_D < 1/3 \)). As the EPC strength increases into the intermediate regime (\( 1/3 < \lambda_D < 1/2 \)), the role of $t^{\prime}$  reverses: It promotes CDW order while suppressing SC pairing. 
This transition is governed by a critical EPC strength corresponding to \( g \approx t \), above which local phonons exert a stronger attractive force on electron pairs, driving the system toward an insulating CDW phase at the expense of SC correlations.

Notably, the effect of $t^{\prime}$  on CDW order depends strongly on electron filling. 
At half filling and fixed EPC strength, \( t' \) suppresses CDW formation and reduces the CDW transition temperature \( T_c \). 
In contrast, near vHS fillings (\( \langle n \rangle = 3/4 \)) in the intermediate EPC regime, $t^{\prime}$  facilitates CDW order, with a CDW gap emerging at a relatively low temperature \( T_c \approx |t|/20 \), which is lower than the CDW transition temperature observed at half filling. 
These results highlight a competitive interplay between SC pairing and CDW order near vHS fillings, where $t^{\prime}$ acts as a tunable knob to favor one order over the other depending on the EPC strength.

Our study provides insights into the phonon-mediated interplay between correlated electronic phases in honeycomb lattice systems, with implications for understanding similar phenomena in some two-dimensional honeycomb structures. 
While our results are derived from this minimal model, they provide generic trends that are expected to hold in systems where an effectively local coupling to optical phonon modes plays a dominant role in the pairing or CDW glue. 
Therefore, we believe that the qualitative trends we identify can be relevant to real materials exhibiting similar structural motifs. 
Future work could extend this investigation to more complex models with more relevance to real materials, including a Holstein model with dispersive phonon modes, as well as optical and bond-centered Su-Schrieffer-Heeger models, to explore the complicated relationship between SC and CDW orders, and potential emergent phases, such as the supersolid and Kekulé valence bond solid phase.

\section*{Data Availability Statement}
The data that support the findings of this study have been deposited in a Zenodo repository \cite{liu_2026_18616136}. 

\appendix

\section{Results of the DEAC method}
\label{app:DEAC}

We perform an analytic continuation using the DEAC method as implemented in \texttt{SmoQyDEAC.jl}~\cite{10.21468/SciPostPhysCodeb.39,10.21468/SciPostPhysCodeb.39-r1.1}.
Whenever binned DQMC data are available, we use \texttt{DEAC\_Binned}, which accounts for correlated noise in imaginary time by diagonalizing the covariance matrix and performing the $\chi^2$ fitting in its eigenbasis.
In addition, bootstrap resampling is employed to ensure a stable covariance estimate with an effective number of bins well exceeding $N_\tau$.
In DEAC, the spectrum is discretized on a fixed frequency grid $\{\omega_j\}$ and represented by a trial spectrum $A_i(\omega_j)$.
The corresponding imaginary-time correlator is obtained from the forward model
\begin{equation}
G_i(\tau_l)=\sum_{j=1}^{N_\omega}\Delta\omega\,K(\tau_l,\omega_j)\,A_i(\omega_j),
\label{eq:deac_forward}
\end{equation}
with the fermionic kernel $K(\tau,\omega)=e^{-\omega\tau}/(1+e^{-\beta\omega})$.
To reduce run-to-run noise, we perform many independent reconstructions and form a fitness-weighted average,
\begin{equation}
A_{\mathrm{bin}}(\omega)=\frac{\sum_{\mu}\chi_{\mu}^{-2}\,A_{\mu}(\omega)}{\sum_{\mu}\chi_{\mu}^{-2}}.
\label{eq:deac_weighted}
\end{equation}

\begin{figure}[ht]
  \centering
  \includegraphics[scale=0.355]{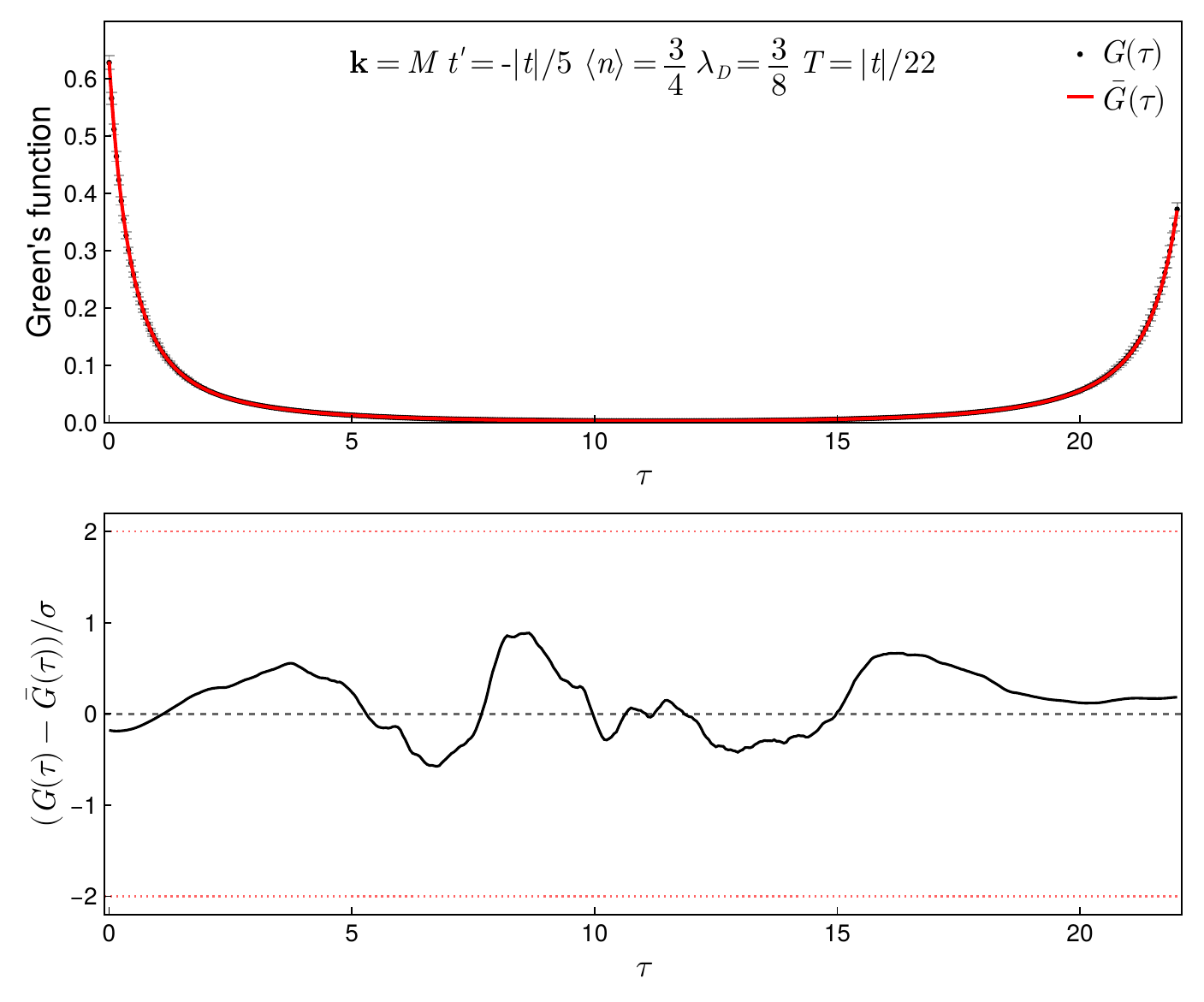}
  \captionsetup{justification=raggedright, singlelinecheck=false}
  \caption{
    Time-dependent Green's function as a function of imaginary time $\tau$, from original DQMC data (black dots) and integrating spectral function (red line), respectively, at $\textbf{k}=M$, $t'=-|t|/5$, $\langle n \rangle = 3/4$, $\lambda_{D}=3/8$, $T=|t|/22$, and $L=6$ (upper). The normalized residual as a function of imaginary time $\tau$ (lower).
  }
  \label{FIG8}
\end{figure}

As a forward-consistency check, Fig.~\ref{FIG8} compares the original time-dependent Green's function $G(\tau)$ from DQMC and the reconstructed $\bar{G}(\tau)=\int d\omega\,K(\tau,\omega)A(\omega)$ at a representative parameter set.
The agreement within error bars and the absence of systematic residuals support the robustness of the spectral functions used in the main text.

\section{Results from adiabatic to antiadiabatic limit}
\label{app:frequency}

\begin{figure}[ht]
  \centering
  \includegraphics[scale=0.22]{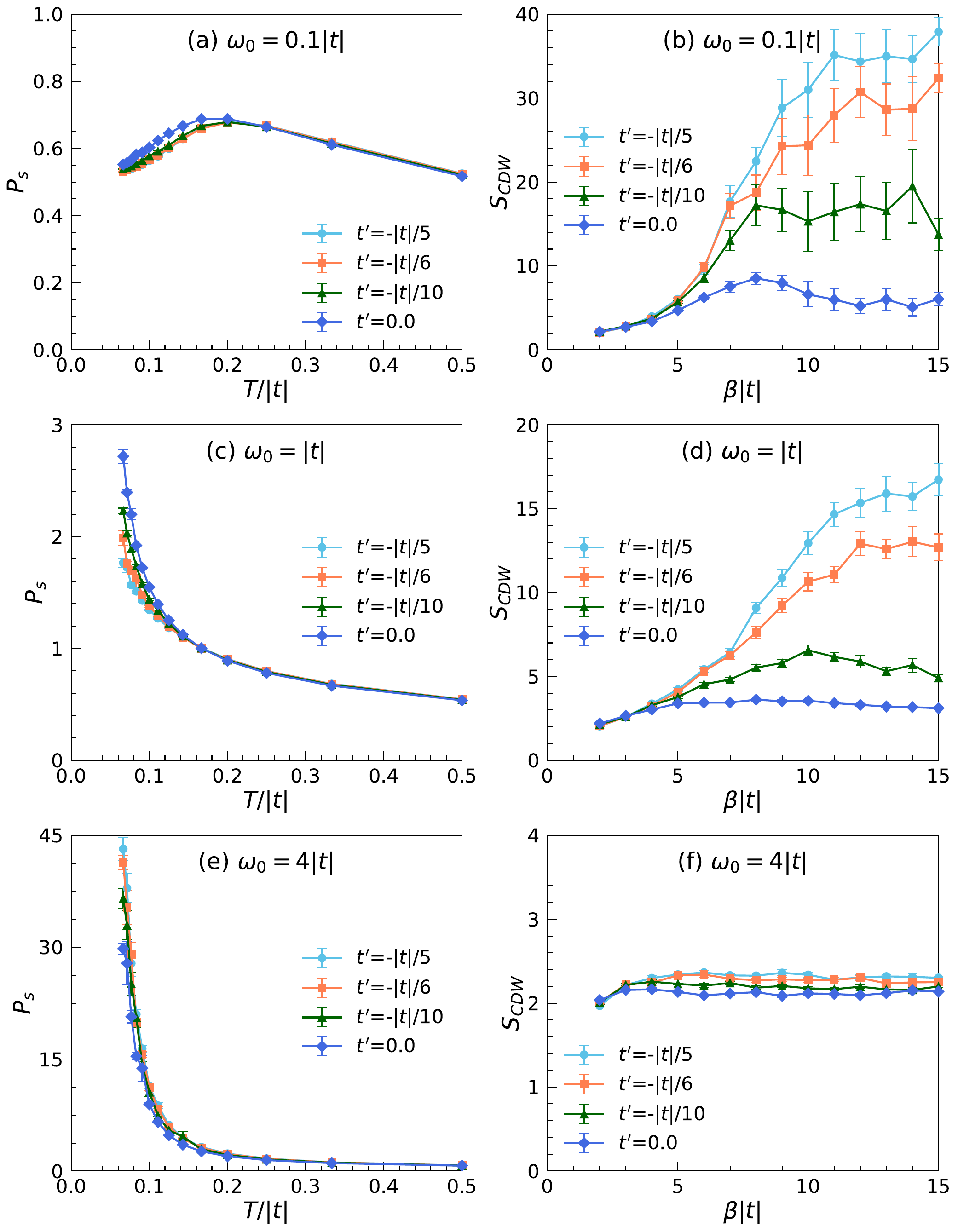}
  \captionsetup{justification=raggedright, singlelinecheck=false}
  \caption{
    Pairing susceptibility $P_{s}$ vs $T/|t|$ (left column) and CDW structure factor $S_{\text{CDW}}$ vs $\beta|t|$ (right column) for several $t'$ at $\lambda_{D}=3/8$, $\langle n\rangle=3/4$, and $L=6$.
    Rows: $\omega_{0}=0.1|t|$, $|t|$, and $4|t|$ (top to bottom).
  }
  \label{FIG9}
\end{figure}

In the main text we choose an intermediate phonon frequency $\omega_{0}=|t|$ to probe a retarded regime where the phonon-mediated interaction is neither quasistatic nor effectively instantaneous.
Figure~\ref{FIG9} compares results at $\omega_{0}=0.1|t|$, $|t|$, and $4|t|$ for fixed $\lambda_{D}=3/8$, $\langle n\rangle=3/4$, and $L=6$.
Over the accessible temperature range, $\omega_{0}=0.1|t|$ shows rapidly enhanced CDW correlations with strongly suppressed pairing, while the trend is reversed at $\omega_{0}=4|t|$.
Therefore, $\omega_{0}=|t|$ provides a balanced window where both tendencies are comparable, making the effect of band-structure tuning via $t'$ on the CDW-SC competition most transparent within our nonperturbative DQMC treatment including vertex corrections.

\bibliography{reference}

\begin{thebibliography}{37}%
\makeatletter
\providecommand \@ifxundefined [1]{%
 \@ifx{#1\undefined}
}%
\providecommand \@ifnum [1]{%
 \ifnum #1\expandafter \@firstoftwo
 \else \expandafter \@secondoftwo
 \fi
}%
\providecommand \@ifx [1]{%
 \ifx #1\expandafter \@firstoftwo
 \else \expandafter \@secondoftwo
 \fi
}%
\providecommand \natexlab [1]{#1}%
\providecommand \enquote  [1]{``#1''}%
\providecommand \bibnamefont  [1]{#1}%
\providecommand \bibfnamefont [1]{#1}%
\providecommand \citenamefont [1]{#1}%
\providecommand \href@noop [0]{\@secondoftwo}%
\providecommand \href [0]{\begingroup \@sanitize@url \@href}%
\providecommand \@href[1]{\@@startlink{#1}\@@href}%
\providecommand \@@href[1]{\endgroup#1\@@endlink}%
\providecommand \@sanitize@url [0]{\catcode `\\12\catcode `\$12\catcode `\&12\catcode `\#12\catcode `\^12\catcode `\_12\catcode `\%12\relax}%
\providecommand \@@startlink[1]{}%
\providecommand \@@endlink[0]{}%
\providecommand \url  [0]{\begingroup\@sanitize@url \@url }%
\providecommand \@url [1]{\endgroup\@href {#1}{\urlprefix }}%
\providecommand \urlprefix  [0]{URL }%
\providecommand \Eprint [0]{\href }%
\providecommand \doibase [0]{https://doi.org/}%
\providecommand \selectlanguage [0]{\@gobble}%
\providecommand \bibinfo  [0]{\@secondoftwo}%
\providecommand \bibfield  [0]{\@secondoftwo}%
\providecommand \translation [1]{[#1]}%
\providecommand \BibitemOpen [0]{}%
\providecommand \bibitemStop [0]{}%
\providecommand \bibitemNoStop [0]{.\EOS\space}%
\providecommand \EOS [0]{\spacefactor3000\relax}%
\providecommand \BibitemShut  [1]{\csname bibitem#1\endcsname}%
\let\auto@bib@innerbib\@empty
\bibitem [{\citenamefont {Merritt}\ \emph {et~al.}(2019)\citenamefont {Merritt}, \citenamefont {Castellan}, \citenamefont {Keller}, \citenamefont {Park}, \citenamefont {Fernandez-Baca}, \citenamefont {Gu},\ and\ \citenamefont {Reznik}}]{merrittLowenergyPhononsBi2019}%
  \BibitemOpen
  \bibfield  {author} {\bibinfo {author} {\bibfnamefont {A.~M.}\ \bibnamefont {Merritt}}, \bibinfo {author} {\bibfnamefont {J.-P.}\ \bibnamefont {Castellan}}, \bibinfo {author} {\bibfnamefont {T.}~\bibnamefont {Keller}}, \bibinfo {author} {\bibfnamefont {S.~R.}\ \bibnamefont {Park}}, \bibinfo {author} {\bibfnamefont {J.~A.}\ \bibnamefont {Fernandez-Baca}}, \bibinfo {author} {\bibfnamefont {G.~D.}\ \bibnamefont {Gu}},\ and\ \bibinfo {author} {\bibfnamefont {D.}~\bibnamefont {Reznik}},\ }\bibfield  {title} {\bibinfo {title} {Low-energy phonons in {Bi}$_{2}${Sr}$_{2}${CaCu}$_{2}${O}$_{8+\delta}$ and their possible interaction with electrons measured by inelastic neutron scattering},\ }\href {https://doi.org/10.1103/PhysRevB.100.144502} {\bibfield  {journal} {\bibinfo  {journal} {Phys. Rev. B}\ }\textbf {\bibinfo {volume} {100}},\ \bibinfo {pages} {144502} (\bibinfo {year} {2019})}\BibitemShut {NoStop}%
\bibitem [{\citenamefont {Pintschovius}(2005)}]{pintschoviusElectronPhononCoupling2005}%
  \BibitemOpen
  \bibfield  {author} {\bibinfo {author} {\bibfnamefont {L.}~\bibnamefont {Pintschovius}},\ }\bibfield  {title} {\bibinfo {title} {Electron–phonon coupling effects explored by inelastic neutron scattering},\ }\href {https://doi.org/10.1002/pssb.200404951} {\bibfield  {journal} {\bibinfo  {journal} {Phys. Status Solidi (b)}\ }\textbf {\bibinfo {volume} {242}},\ \bibinfo {pages} {30} (\bibinfo {year} {2005})}\BibitemShut {NoStop}%
\bibitem [{\citenamefont {Ideta}\ \emph {et~al.}(2013)\citenamefont {Ideta}, \citenamefont {Yoshida}, \citenamefont {Hashimoto}, \citenamefont {Fujimori}, \citenamefont {Anzai}, \citenamefont {Ino}, \citenamefont {Arita}, \citenamefont {Namatame}, \citenamefont {Taniguchi}, \citenamefont {Takashima}, \citenamefont {Kojima},\ and\ \citenamefont {Uchida}}]{idetaEffectElectronphononCoupling2013}%
  \BibitemOpen
  \bibfield  {author} {\bibinfo {author} {\bibfnamefont {S.}~\bibnamefont {Ideta}}, \bibinfo {author} {\bibfnamefont {T.}~\bibnamefont {Yoshida}}, \bibinfo {author} {\bibfnamefont {M.}~\bibnamefont {Hashimoto}}, \bibinfo {author} {\bibfnamefont {A.}~\bibnamefont {Fujimori}}, \bibinfo {author} {\bibfnamefont {H.}~\bibnamefont {Anzai}}, \bibinfo {author} {\bibfnamefont {A.}~\bibnamefont {Ino}}, \bibinfo {author} {\bibfnamefont {M.}~\bibnamefont {Arita}}, \bibinfo {author} {\bibfnamefont {H.}~\bibnamefont {Namatame}}, \bibinfo {author} {\bibfnamefont {M.}~\bibnamefont {Taniguchi}}, \bibinfo {author} {\bibfnamefont {K.}~\bibnamefont {Takashima}}, \bibinfo {author} {\bibfnamefont {K.~M.}\ \bibnamefont {Kojima}},\ and\ \bibinfo {author} {\bibfnamefont {S.}~\bibnamefont {Uchida}},\ }\bibfield  {title} {\bibinfo {title} {Effect of electron-phonon coupling in the {ARPES} spectra of the tri-layer cuprate {Bi}$_{2}${Sr}$_{2}${Ca}$_{2}${Cu}$_{3}${O}$_{10+\delta}$},\ }\href {https://doi.org/10.1088/1742-6596/428/1/012039} {\bibfield  {journal} {\bibinfo  {journal} {J. Phys.: Conf. Ser.}\ }\textbf {\bibinfo {volume} {428}},\ \bibinfo {pages} {012039} (\bibinfo {year} {2013})}\BibitemShut {NoStop}%
\bibitem [{\citenamefont {He}\ \emph {et~al.}(2018)\citenamefont {He}, \citenamefont {Hashimoto}, \citenamefont {Song}, \citenamefont {Chen}, \citenamefont {He}, \citenamefont {Vishik}, \citenamefont {Moritz}, \citenamefont {Lee}, \citenamefont {Nagaosa}, \citenamefont {Zaanen}, \citenamefont {Devereaux}, \citenamefont {Yoshida}, \citenamefont {Eisaki}, \citenamefont {Lu},\ and\ \citenamefont {Shen}}]{heRapidChangeSuperconductivity2018}%
  \BibitemOpen
  \bibfield  {author} {\bibinfo {author} {\bibfnamefont {Y.}~\bibnamefont {He}}, \bibinfo {author} {\bibfnamefont {M.}~\bibnamefont {Hashimoto}}, \bibinfo {author} {\bibfnamefont {D.}~\bibnamefont {Song}}, \bibinfo {author} {\bibfnamefont {S.-D.}\ \bibnamefont {Chen}}, \bibinfo {author} {\bibfnamefont {J.}~\bibnamefont {He}}, \bibinfo {author} {\bibfnamefont {I.~M.}\ \bibnamefont {Vishik}}, \bibinfo {author} {\bibfnamefont {B.}~\bibnamefont {Moritz}}, \bibinfo {author} {\bibfnamefont {D.-H.}\ \bibnamefont {Lee}}, \bibinfo {author} {\bibfnamefont {N.}~\bibnamefont {Nagaosa}}, \bibinfo {author} {\bibfnamefont {J.}~\bibnamefont {Zaanen}}, \bibinfo {author} {\bibfnamefont {T.~P.}\ \bibnamefont {Devereaux}}, \bibinfo {author} {\bibfnamefont {Y.}~\bibnamefont {Yoshida}}, \bibinfo {author} {\bibfnamefont {H.}~\bibnamefont {Eisaki}}, \bibinfo {author} {\bibfnamefont {D.~H.}\ \bibnamefont {Lu}},\ and\ \bibinfo {author} {\bibfnamefont {Z.-X.}\ \bibnamefont {Shen}},\ }\bibfield  {title} {\bibinfo {title} {Rapid change of superconductivity and electron-phonon coupling through critical doping in {{Bi-2212}}},\ }\href {https://doi.org/10.1126/science.aar3394} {\bibfield  {journal} {\bibinfo  {journal} {Science}\ }\textbf {\bibinfo {volume} {362}},\ \bibinfo {pages} {62} (\bibinfo {year} {2018})}\BibitemShut {NoStop}%
\bibitem [{\citenamefont {Kendziora}\ \emph {et~al.}(2001)\citenamefont {Kendziora}, \citenamefont {Nachumi}, \citenamefont {Fournier}, \citenamefont {Li}, \citenamefont {Greene},\ and\ \citenamefont {Hinks}}]{kendzioraUnconventionalSuperconductivityObserved2001}%
  \BibitemOpen
  \bibfield  {author} {\bibinfo {author} {\bibfnamefont {C.}~\bibnamefont {Kendziora}}, \bibinfo {author} {\bibfnamefont {B.}~\bibnamefont {Nachumi}}, \bibinfo {author} {\bibfnamefont {P.}~\bibnamefont {Fournier}}, \bibinfo {author} {\bibfnamefont {Z.}~\bibnamefont {Li}}, \bibinfo {author} {\bibfnamefont {R.}~\bibnamefont {Greene}},\ and\ \bibinfo {author} {\bibfnamefont {D.}~\bibnamefont {Hinks}},\ }\bibfield  {title} {\bibinfo {title} {Unconventional superconductivity observed with raman spectroscopy in p- and n-type cuprates},\ }\href {https://doi.org/10.1016/S0921-4534(01)00847-4} {\bibfield  {journal} {\bibinfo  {journal} {Physica C}\ }\textbf {\bibinfo {volume} {364--365}},\ \bibinfo {pages} {541} (\bibinfo {year} {2001})}\BibitemShut {NoStop}%
\bibitem [{\citenamefont {Luo}\ \emph {et~al.}(2022)\citenamefont {Luo}, \citenamefont {Gao}, \citenamefont {Liu}, \citenamefont {Gu}, \citenamefont {Wu}, \citenamefont {Yi}, \citenamefont {Jia}, \citenamefont {Wu}, \citenamefont {Luo}, \citenamefont {Xu}, \citenamefont {Zhao}, \citenamefont {Wang}, \citenamefont {Mao}, \citenamefont {Liu}, \citenamefont {Zhu}, \citenamefont {Shi}, \citenamefont {Jiang}, \citenamefont {Hu}, \citenamefont {Xu},\ and\ \citenamefont {Zhou}}]{luoElectronicNatureCharge2022}%
  \BibitemOpen
  \bibfield  {author} {\bibinfo {author} {\bibfnamefont {H.}~\bibnamefont {Luo}}, \bibinfo {author} {\bibfnamefont {Q.}~\bibnamefont {Gao}}, \bibinfo {author} {\bibfnamefont {H.}~\bibnamefont {Liu}}, \bibinfo {author} {\bibfnamefont {Y.}~\bibnamefont {Gu}}, \bibinfo {author} {\bibfnamefont {D.}~\bibnamefont {Wu}}, \bibinfo {author} {\bibfnamefont {C.}~\bibnamefont {Yi}}, \bibinfo {author} {\bibfnamefont {J.}~\bibnamefont {Jia}}, \bibinfo {author} {\bibfnamefont {S.}~\bibnamefont {Wu}}, \bibinfo {author} {\bibfnamefont {X.}~\bibnamefont {Luo}}, \bibinfo {author} {\bibfnamefont {Y.}~\bibnamefont {Xu}}, \bibinfo {author} {\bibfnamefont {L.}~\bibnamefont {Zhao}}, \bibinfo {author} {\bibfnamefont {Q.}~\bibnamefont {Wang}}, \bibinfo {author} {\bibfnamefont {H.}~\bibnamefont {Mao}}, \bibinfo {author} {\bibfnamefont {G.}~\bibnamefont {Liu}}, \bibinfo {author} {\bibfnamefont {Z.}~\bibnamefont {Zhu}}, \bibinfo {author} {\bibfnamefont {Y.}~\bibnamefont {Shi}}, \bibinfo {author} {\bibfnamefont {K.}~\bibnamefont {Jiang}}, \bibinfo {author} {\bibfnamefont {J.}~\bibnamefont {Hu}}, \bibinfo {author} {\bibfnamefont {Z.}~\bibnamefont {Xu}},\ and\ \bibinfo {author} {\bibfnamefont {X.~J.}\ \bibnamefont {Zhou}},\ }\bibfield  {title} {\bibinfo {title} {Electronic nature of charge density wave and electron-phonon coupling in kagome superconductor {K}{V}$_{2}${Sb}$_{5}$},\ }\href {https://doi.org/10.1038/s41467-021-27946-6} {\bibfield  {journal} {\bibinfo  {journal} {Nat. Commun.}\ }\textbf {\bibinfo {volume} {13}},\ \bibinfo {pages} {273} (\bibinfo {year} {2022})}\BibitemShut {NoStop}%
\bibitem [{\citenamefont {Ang}\ \emph {et~al.}(2012)\citenamefont {Ang}, \citenamefont {Tanaka}, \citenamefont {Ieki}, \citenamefont {Nakayama}, \citenamefont {Sato}, \citenamefont {Li}, \citenamefont {Lu}, \citenamefont {Sun},\ and\ \citenamefont {Takahashi}}]{angRealspaceCoexistenceMelted2012}%
  \BibitemOpen
  \bibfield  {author} {\bibinfo {author} {\bibfnamefont {R.}~\bibnamefont {Ang}}, \bibinfo {author} {\bibfnamefont {Y.}~\bibnamefont {Tanaka}}, \bibinfo {author} {\bibfnamefont {E.}~\bibnamefont {Ieki}}, \bibinfo {author} {\bibfnamefont {K.}~\bibnamefont {Nakayama}}, \bibinfo {author} {\bibfnamefont {T.}~\bibnamefont {Sato}}, \bibinfo {author} {\bibfnamefont {L.~J.}\ \bibnamefont {Li}}, \bibinfo {author} {\bibfnamefont {W.~J.}\ \bibnamefont {Lu}}, \bibinfo {author} {\bibfnamefont {Y.~P.}\ \bibnamefont {Sun}},\ and\ \bibinfo {author} {\bibfnamefont {T.}~\bibnamefont {Takahashi}},\ }\bibfield  {title} {\bibinfo {title} {Real-space coexistence of the melted mott state and superconductivity in {Fe-substituted}1$\textit{T}$-{TaS}$_{2}$},\ }\href {https://doi.org/10.1103/PhysRevLett.109.176403} {\bibfield  {journal} {\bibinfo  {journal} {Phys. Rev. Lett.}\ }\textbf {\bibinfo {volume} {109}},\ \bibinfo {pages} {176403} (\bibinfo {year} {2012})}\BibitemShut {NoStop}%
\bibitem [{\citenamefont {Liu}\ \emph {et~al.}(2013)\citenamefont {Liu}, \citenamefont {Ang}, \citenamefont {Lu}, \citenamefont {Song}, \citenamefont {Li},\ and\ \citenamefont {Sun}}]{liuSuperconductivityInducedSedoping2013}%
  \BibitemOpen
  \bibfield  {author} {\bibinfo {author} {\bibfnamefont {Y.}~\bibnamefont {Liu}}, \bibinfo {author} {\bibfnamefont {R.}~\bibnamefont {Ang}}, \bibinfo {author} {\bibfnamefont {W.~J.}\ \bibnamefont {Lu}}, \bibinfo {author} {\bibfnamefont {W.~H.}\ \bibnamefont {Song}}, \bibinfo {author} {\bibfnamefont {L.~J.}\ \bibnamefont {Li}},\ and\ \bibinfo {author} {\bibfnamefont {Y.~P.}\ \bibnamefont {Sun}},\ }\bibfield  {title} {\bibinfo {title} {Superconductivity induced by {{Se-doping}} in layered charge-density-wave system 1$\textit{T}$-{Ta}{S}$_{2-x}${Se}$_{x}$},\ }\href {https://doi.org/10.1063/1.4805003} {\bibfield  {journal} {\bibinfo  {journal} {Appl. Phys. Lett.}\ }\textbf {\bibinfo {volume} {102}},\ \bibinfo {pages} {192602} (\bibinfo {year} {2013})}\BibitemShut {NoStop}%
\bibitem [{\citenamefont {Cho}\ \emph {et~al.}(2018)\citenamefont {Cho}, \citenamefont {Kończykowski}, \citenamefont {Teknowijoyo}, \citenamefont {Tanatar}, \citenamefont {Guss}, \citenamefont {Gartin}, \citenamefont {Wilde}, \citenamefont {Kreyssig}, \citenamefont {McQueeney}, \citenamefont {Goldman}, \citenamefont {Mishra}, \citenamefont {Hirschfeld},\ and\ \citenamefont {Prozorov}}]{choUsingControlledDisorder2018}%
  \BibitemOpen
  \bibfield  {author} {\bibinfo {author} {\bibfnamefont {K.}~\bibnamefont {Cho}}, \bibinfo {author} {\bibfnamefont {M.}~\bibnamefont {Kończykowski}}, \bibinfo {author} {\bibfnamefont {S.}~\bibnamefont {Teknowijoyo}}, \bibinfo {author} {\bibfnamefont {M.~A.}\ \bibnamefont {Tanatar}}, \bibinfo {author} {\bibfnamefont {J.}~\bibnamefont {Guss}}, \bibinfo {author} {\bibfnamefont {P.~B.}\ \bibnamefont {Gartin}}, \bibinfo {author} {\bibfnamefont {J.~M.}\ \bibnamefont {Wilde}}, \bibinfo {author} {\bibfnamefont {A.}~\bibnamefont {Kreyssig}}, \bibinfo {author} {\bibfnamefont {R.~J.}\ \bibnamefont {McQueeney}}, \bibinfo {author} {\bibfnamefont {A.~I.}\ \bibnamefont {Goldman}}, \bibinfo {author} {\bibfnamefont {V.}~\bibnamefont {Mishra}}, \bibinfo {author} {\bibfnamefont {P.~J.}\ \bibnamefont {Hirschfeld}},\ and\ \bibinfo {author} {\bibfnamefont {R.}~\bibnamefont {Prozorov}},\ }\bibfield  {title} {\bibinfo {title} {Using controlled disorder to probe the interplay between charge order and superconductivity in {Nb}{Se}$_{2}$},\ }\href {https://doi.org/10.1038/s41467-018-05153-0} {\bibfield  {journal} {\bibinfo  {journal} {Nat. Commun.}\ }\textbf {\bibinfo {volume} {9}},\ \bibinfo {pages} {2796} (\bibinfo {year} {2018})}\BibitemShut {NoStop}%
\bibitem [{\citenamefont {Bardeen}\ \emph {et~al.}(1957)\citenamefont {Bardeen}, \citenamefont {Cooper},\ and\ \citenamefont {Schrieffer}}]{bardeenTheorySuperconductivity1957}%
  \BibitemOpen
  \bibfield  {author} {\bibinfo {author} {\bibfnamefont {J.}~\bibnamefont {Bardeen}}, \bibinfo {author} {\bibfnamefont {L.~N.}\ \bibnamefont {Cooper}},\ and\ \bibinfo {author} {\bibfnamefont {J.~R.}\ \bibnamefont {Schrieffer}},\ }\bibfield  {title} {\bibinfo {title} {Theory of {{Superconductivity}}},\ }\href {https://doi.org/10.1103/PhysRev.108.1175} {\bibfield  {journal} {\bibinfo  {journal} {Phys. Rev.}\ }\textbf {\bibinfo {volume} {108}},\ \bibinfo {pages} {1175} (\bibinfo {year} {1957})}\BibitemShut {NoStop}%
\bibitem [{\citenamefont {Zhu}\ \emph {et~al.}(2015)\citenamefont {Zhu}, \citenamefont {Cao}, \citenamefont {Zhang}, \citenamefont {Plummer},\ and\ \citenamefont {Guo}}]{zhuClassificationChargeDensity2015}%
  \BibitemOpen
  \bibfield  {author} {\bibinfo {author} {\bibfnamefont {X.}~\bibnamefont {Zhu}}, \bibinfo {author} {\bibfnamefont {Y.}~\bibnamefont {Cao}}, \bibinfo {author} {\bibfnamefont {J.}~\bibnamefont {Zhang}}, \bibinfo {author} {\bibfnamefont {E.~W.}\ \bibnamefont {Plummer}},\ and\ \bibinfo {author} {\bibfnamefont {J.}~\bibnamefont {Guo}},\ }\bibfield  {title} {\bibinfo {title} {Classification of charge density waves based on their nature},\ }\href {https://doi.org/10.1073/pnas.1424791112} {\bibfield  {journal} {\bibinfo  {journal} {Proc. Natl. Acad. Sci. U.S.A.}\ }\textbf {\bibinfo {volume} {112}},\ \bibinfo {pages} {2367} (\bibinfo {year} {2015})}\BibitemShut {NoStop}%
\bibitem [{\citenamefont {Grüner}(1988)}]{grunerDynamicsChargedensityWaves1988}%
  \BibitemOpen
  \bibfield  {author} {\bibinfo {author} {\bibfnamefont {G.}~\bibnamefont {Grüner}},\ }\bibfield  {title} {\bibinfo {title} {The dynamics of charge-density waves},\ }\href {https://doi.org/10.1103/RevModPhys.60.1129} {\bibfield  {journal} {\bibinfo  {journal} {Rev. Mod. Phys.}\ }\textbf {\bibinfo {volume} {60}},\ \bibinfo {pages} {1129} (\bibinfo {year} {1988})}\BibitemShut {NoStop}%
\bibitem [{\citenamefont {Lin}\ \emph {et~al.}(2015)\citenamefont {Lin}, \citenamefont {Ma},\ and\ \citenamefont {Huang}}]{linQuantumMonteCarlo2015}%
  \BibitemOpen
  \bibfield  {author} {\bibinfo {author} {\bibfnamefont {H.}~\bibnamefont {Lin}}, \bibinfo {author} {\bibfnamefont {T.}~\bibnamefont {Ma}},\ and\ \bibinfo {author} {\bibfnamefont {Z.}~\bibnamefont {Huang}},\ }\bibfield  {title} {\bibinfo {title} {Quantum {{Monte Carlo}} study of magnetic and superconducting properties of graphene},\ }\href {https://doi.org/10.1002/mma.2851} {\bibfield  {journal} {\bibinfo  {journal} {Math Methods in App Sciences}\ }\textbf {\bibinfo {volume} {38}},\ \bibinfo {pages} {4487} (\bibinfo {year} {2015})}\BibitemShut {NoStop}%
\bibitem [{\citenamefont {Holstein}(2000)}]{WOS:000086983000016}%
  \BibitemOpen
  \bibfield  {author} {\bibinfo {author} {\bibfnamefont {T.}~\bibnamefont {Holstein}},\ }\bibfield  {title} {\bibinfo {title} {Studies of polaron motion: Part {I}. {The} molecular-crystal model},\ }\href {https://doi.org/10.1006/aphy.2000.6020} {\bibfield  {journal} {\bibinfo  {journal} {ANNALS OF PHYSICS}\ }\textbf {\bibinfo {volume} {281}},\ \bibinfo {pages} {706} (\bibinfo {year} {2000})}\BibitemShut {NoStop}%
\bibitem [{\citenamefont {Zhang}\ \emph {et~al.}(2019)\citenamefont {Zhang}, \citenamefont {Chiu}, \citenamefont {Costa}, \citenamefont {Batrouni},\ and\ \citenamefont {Scalettar}}]{zhangChargeOrderHolstein2019}%
  \BibitemOpen
  \bibfield  {author} {\bibinfo {author} {\bibfnamefont {Y.-X.}\ \bibnamefont {Zhang}}, \bibinfo {author} {\bibfnamefont {W.-T.}\ \bibnamefont {Chiu}}, \bibinfo {author} {\bibfnamefont {N.~C.}\ \bibnamefont {Costa}}, \bibinfo {author} {\bibfnamefont {G.~G.}\ \bibnamefont {Batrouni}},\ and\ \bibinfo {author} {\bibfnamefont {R.~T.}\ \bibnamefont {Scalettar}},\ }\bibfield  {title} {\bibinfo {title} {Charge {{Order}} in the {{Holstein Model}} on a {{Honeycomb Lattice}}},\ }\href {https://doi.org/10.1103/PhysRevLett.122.077602} {\bibfield  {journal} {\bibinfo  {journal} {Phys. Rev. Lett.}\ }\textbf {\bibinfo {volume} {122}},\ \bibinfo {pages} {077602} (\bibinfo {year} {2019})}\BibitemShut {NoStop}%
\bibitem [{\citenamefont {Feng}\ and\ \citenamefont {Scalettar}(2020)}]{fengInterplayFlatElectronic2020}%
  \BibitemOpen
  \bibfield  {author} {\bibinfo {author} {\bibfnamefont {C.}~\bibnamefont {Feng}}\ and\ \bibinfo {author} {\bibfnamefont {R.~T.}\ \bibnamefont {Scalettar}},\ }\bibfield  {title} {\bibinfo {title} {Interplay of flat electronic bands with {{Holstein}} phonons},\ }\href {https://doi.org/10.1103/PhysRevB.102.235152} {\bibfield  {journal} {\bibinfo  {journal} {Phys. Rev. B}\ }\textbf {\bibinfo {volume} {102}},\ \bibinfo {pages} {235152} (\bibinfo {year} {2020})}\BibitemShut {NoStop}%
\bibitem [{\citenamefont {Nosarzewski}\ \emph {et~al.}(2021)\citenamefont {Nosarzewski}, \citenamefont {Huang}, \citenamefont {Dee}, \citenamefont {Esterlis}, \citenamefont {Moritz}, \citenamefont {Kivelson}, \citenamefont {Johnston},\ and\ \citenamefont {Devereaux}}]{nosarzewskiSuperconductivityChargeDensity2021}%
  \BibitemOpen
  \bibfield  {author} {\bibinfo {author} {\bibfnamefont {B.}~\bibnamefont {Nosarzewski}}, \bibinfo {author} {\bibfnamefont {E.~W.}\ \bibnamefont {Huang}}, \bibinfo {author} {\bibfnamefont {P.~M.}\ \bibnamefont {Dee}}, \bibinfo {author} {\bibfnamefont {I.}~\bibnamefont {Esterlis}}, \bibinfo {author} {\bibfnamefont {B.}~\bibnamefont {Moritz}}, \bibinfo {author} {\bibfnamefont {S.~A.}\ \bibnamefont {Kivelson}}, \bibinfo {author} {\bibfnamefont {S.}~\bibnamefont {Johnston}},\ and\ \bibinfo {author} {\bibfnamefont {T.~P.}\ \bibnamefont {Devereaux}},\ }\bibfield  {title} {\bibinfo {title} {Superconductivity, charge density waves, and bipolarons in the {{Holstein}} model},\ }\href {https://doi.org/10.1103/PhysRevB.103.235156} {\bibfield  {journal} {\bibinfo  {journal} {Phys. Rev. B}\ }\textbf {\bibinfo {volume} {103}},\ \bibinfo {pages} {235156} (\bibinfo {year} {2021})}\BibitemShut {NoStop}%
\bibitem [{\citenamefont {Li}\ \emph {et~al.}(2019)\citenamefont {Li}, \citenamefont {Cohen},\ and\ \citenamefont {Lee}}]{liEnhancementSuperconductivityFrustrating2019}%
  \BibitemOpen
  \bibfield  {author} {\bibinfo {author} {\bibfnamefont {Z.-X.}\ \bibnamefont {Li}}, \bibinfo {author} {\bibfnamefont {M.~L.}\ \bibnamefont {Cohen}},\ and\ \bibinfo {author} {\bibfnamefont {D.-H.}\ \bibnamefont {Lee}},\ }\bibfield  {title} {\bibinfo {title} {Enhancement of superconductivity by frustrating the charge order},\ }\href {https://doi.org/10.1103/PhysRevB.100.245105} {\bibfield  {journal} {\bibinfo  {journal} {Phys. Rev. B}\ }\textbf {\bibinfo {volume} {100}},\ \bibinfo {pages} {245105} (\bibinfo {year} {2019})}\BibitemShut {NoStop}%
\bibitem [{\citenamefont {Bradley}\ \emph {et~al.}(2023)\citenamefont {Bradley}, \citenamefont {Cohen-Stead}, \citenamefont {Johnston}, \citenamefont {Barros},\ and\ \citenamefont {Scalettar}}]{bradleyChargeOrderKagome2023}%
  \BibitemOpen
  \bibfield  {author} {\bibinfo {author} {\bibfnamefont {O.}~\bibnamefont {Bradley}}, \bibinfo {author} {\bibfnamefont {B.}~\bibnamefont {Cohen-Stead}}, \bibinfo {author} {\bibfnamefont {S.}~\bibnamefont {Johnston}}, \bibinfo {author} {\bibfnamefont {K.}~\bibnamefont {Barros}},\ and\ \bibinfo {author} {\bibfnamefont {R.~T.}\ \bibnamefont {Scalettar}},\ }\bibfield  {title} {\bibinfo {title} {Charge order in the kagome lattice {{Holstein}} model: A hybrid {{Monte Carlo}} study},\ }\href {https://doi.org/10.1038/s41535-023-00553-y} {\bibfield  {journal} {\bibinfo  {journal} {npj Quantum Mater.}\ }\textbf {\bibinfo {volume} {8}},\ \bibinfo {pages} {21} (\bibinfo {year} {2023})}\BibitemShut {NoStop}%
\bibitem [{\citenamefont {Costa}\ \emph {et~al.}(2018)\citenamefont {Costa}, \citenamefont {Blommel}, \citenamefont {Chiu}, \citenamefont {Batrouni},\ and\ \citenamefont {Scalettar}}]{costaPhononDispersionCompetition2018}%
  \BibitemOpen
  \bibfield  {author} {\bibinfo {author} {\bibfnamefont {N.~C.}\ \bibnamefont {Costa}}, \bibinfo {author} {\bibfnamefont {T.}~\bibnamefont {Blommel}}, \bibinfo {author} {\bibfnamefont {W.-T.}\ \bibnamefont {Chiu}}, \bibinfo {author} {\bibfnamefont {G.}~\bibnamefont {Batrouni}},\ and\ \bibinfo {author} {\bibfnamefont {R.~T.}\ \bibnamefont {Scalettar}},\ }\bibfield  {title} {\bibinfo {title} {Phonon dispersion and the competition between pairing and charge order},\ }\href {https://doi.org/10.1103/PhysRevLett.120.187003} {\bibfield  {journal} {\bibinfo  {journal} {Phys. Rev. Lett.}\ }\textbf {\bibinfo {volume} {120}},\ \bibinfo {pages} {187003} (\bibinfo {year} {2018})}\BibitemShut {NoStop}%
\bibitem [{\citenamefont {Meng}\ \emph {et~al.}(2024)\citenamefont {Meng}, \citenamefont {Zhang}, \citenamefont {Fernandes}, \citenamefont {Ma},\ and\ \citenamefont {Scalettar}}]{mengSupersolidPhaseDiluted2024}%
  \BibitemOpen
  \bibfield  {author} {\bibinfo {author} {\bibfnamefont {J.}~\bibnamefont {Meng}}, \bibinfo {author} {\bibfnamefont {Y.}~\bibnamefont {Zhang}}, \bibinfo {author} {\bibfnamefont {R.~M.}\ \bibnamefont {Fernandes}}, \bibinfo {author} {\bibfnamefont {T.}~\bibnamefont {Ma}},\ and\ \bibinfo {author} {\bibfnamefont {R.~T.}\ \bibnamefont {Scalettar}},\ }\bibfield  {title} {\bibinfo {title} {Supersolid phase in the diluted {{Holstein}} model},\ }\href {https://doi.org/10.1103/PhysRevB.110.L220506} {\bibfield  {journal} {\bibinfo  {journal} {Phys. Rev. B}\ }\textbf {\bibinfo {volume} {110}},\ \bibinfo {pages} {L220506} (\bibinfo {year} {2024})}\BibitemShut {NoStop}%
\bibitem [{\citenamefont {Ying}\ \emph {et~al.}(2024)\citenamefont {Ying}, \citenamefont {Xu},\ and\ \citenamefont {Guo}}]{yingChargeDensityWave2024}%
  \BibitemOpen
  \bibfield  {author} {\bibinfo {author} {\bibfnamefont {T.}~\bibnamefont {Ying}}, \bibinfo {author} {\bibfnamefont {Y.}~\bibnamefont {Xu}},\ and\ \bibinfo {author} {\bibfnamefont {H.}~\bibnamefont {Guo}},\ }\bibfield  {title} {\bibinfo {title} {Charge density wave and pairing order in the {{Holstein}} model on the honeycomb lattice away from half-filling},\ }\href {https://doi.org/10.1103/PhysRevB.110.205145} {\bibfield  {journal} {\bibinfo  {journal} {Phys. Rev. B}\ }\textbf {\bibinfo {volume} {110}},\ \bibinfo {pages} {205145} (\bibinfo {year} {2024})}\BibitemShut {NoStop}%
\bibitem [{\citenamefont {Xiao}\ \emph {et~al.}(2021)\citenamefont {Xiao}, \citenamefont {Costa}, \citenamefont {Khatami}, \citenamefont {Batrouni},\ and\ \citenamefont {Scalettar}}]{xiaoChargeDensityWave2021}%
  \BibitemOpen
  \bibfield  {author} {\bibinfo {author} {\bibfnamefont {B.}~\bibnamefont {Xiao}}, \bibinfo {author} {\bibfnamefont {N.~C.}\ \bibnamefont {Costa}}, \bibinfo {author} {\bibfnamefont {E.}~\bibnamefont {Khatami}}, \bibinfo {author} {\bibfnamefont {G.~G.}\ \bibnamefont {Batrouni}},\ and\ \bibinfo {author} {\bibfnamefont {R.~T.}\ \bibnamefont {Scalettar}},\ }\bibfield  {title} {\bibinfo {title} {Charge density wave and superconductivity in the disordered {{Holstein}} model},\ }\href {https://doi.org/10.1103/PhysRevB.103.L060501} {\bibfield  {journal} {\bibinfo  {journal} {Phys. Rev. B}\ }\textbf {\bibinfo {volume} {103}},\ \bibinfo {pages} {L060501} (\bibinfo {year} {2021})}\BibitemShut {NoStop}%
\bibitem [{\citenamefont {Vekić}\ \emph {et~al.}(1992)\citenamefont {Vekić}, \citenamefont {Noack},\ and\ \citenamefont {White}}]{vekicChargedensityWavesSuperconductivity1992}%
  \BibitemOpen
  \bibfield  {author} {\bibinfo {author} {\bibfnamefont {M.}~\bibnamefont {Vekić}}, \bibinfo {author} {\bibfnamefont {R.~M.}\ \bibnamefont {Noack}},\ and\ \bibinfo {author} {\bibfnamefont {S.~R.}\ \bibnamefont {White}},\ }\bibfield  {title} {\bibinfo {title} {Charge-density waves versus superconductivity in the {{Holstein}} model with next-nearest-neighbor hopping},\ }\href {https://doi.org/10.1103/PhysRevB.46.271} {\bibfield  {journal} {\bibinfo  {journal} {Phys. Rev. B}\ }\textbf {\bibinfo {volume} {46}},\ \bibinfo {pages} {271} (\bibinfo {year} {1992})}\BibitemShut {NoStop}%
\bibitem [{\citenamefont {Novoselov}\ \emph {et~al.}(2024)\citenamefont {Novoselov}, \citenamefont {Geim}, \citenamefont {Morozov}, \citenamefont {Jiang}, \citenamefont {Zhang}, \citenamefont {Dubonos}, \citenamefont {Grigorieva},\ and\ \citenamefont {Firsov}}]{novoselovElectricFieldEffect2004}%
  \BibitemOpen
  \bibfield  {author} {\bibinfo {author} {\bibfnamefont {K.~S.}\ \bibnamefont {Novoselov}}, \bibinfo {author} {\bibfnamefont {A.~K.}\ \bibnamefont {Geim}}, \bibinfo {author} {\bibfnamefont {S.~V.}\ \bibnamefont {Morozov}}, \bibinfo {author} {\bibfnamefont {D.}~\bibnamefont {Jiang}}, \bibinfo {author} {\bibfnamefont {Y.}~\bibnamefont {Zhang}}, \bibinfo {author} {\bibfnamefont {S.~V.}\ \bibnamefont {Dubonos}}, \bibinfo {author} {\bibfnamefont {I.~V.}\ \bibnamefont {Grigorieva}},\ and\ \bibinfo {author} {\bibfnamefont {A.~A.}\ \bibnamefont {Firsov}},\ }\bibfield  {title} {\bibinfo {title} {Electric {{Field Effect}} in {{Atomically Thin Carbon Films}}},\ }\href {https://doi.org/10.1126/science.1102896} {\bibfield  {journal} {\bibinfo  {journal} {Science}\ }\textbf {\bibinfo {volume} {306}},\ \bibinfo {pages} {666} (\bibinfo {year} {2024})}\BibitemShut {NoStop}%
\bibitem [{\citenamefont {Ma}\ \emph {et~al.}(2010)\citenamefont {Ma}, \citenamefont {Hu}, \citenamefont {Huang},\ and\ \citenamefont {Lin}}]{maControllabilityFerromagnetismGraphene2010}%
  \BibitemOpen
  \bibfield  {author} {\bibinfo {author} {\bibfnamefont {T.}~\bibnamefont {Ma}}, \bibinfo {author} {\bibfnamefont {F.}~\bibnamefont {Hu}}, \bibinfo {author} {\bibfnamefont {Z.}~\bibnamefont {Huang}},\ and\ \bibinfo {author} {\bibfnamefont {H.-Q.}\ \bibnamefont {Lin}},\ }\bibfield  {title} {\bibinfo {title} {Controllability of ferromagnetism in graphene},\ }\href {https://doi.org/10.1063/1.3485059} {\bibfield  {journal} {\bibinfo  {journal} {Appl. Phys. Lett.}\ }\textbf {\bibinfo {volume} {97}},\ \bibinfo {pages} {112504} (\bibinfo {year} {2010})}\BibitemShut {NoStop}%
\bibitem [{\citenamefont {Jia}\ \emph {et~al.}(2022)\citenamefont {Jia}, \citenamefont {Yang}, \citenamefont {Li}, \citenamefont {Yang}, \citenamefont {Ying}, \citenamefont {Li},\ and\ \citenamefont {Sun}}]{jiaPairingHubbardModel2022}%
  \BibitemOpen
  \bibfield  {author} {\bibinfo {author} {\bibfnamefont {P.}~\bibnamefont {Jia}}, \bibinfo {author} {\bibfnamefont {S.}~\bibnamefont {Yang}}, \bibinfo {author} {\bibfnamefont {W.}~\bibnamefont {Li}}, \bibinfo {author} {\bibfnamefont {J.}~\bibnamefont {Yang}}, \bibinfo {author} {\bibfnamefont {T.}~\bibnamefont {Ying}}, \bibinfo {author} {\bibfnamefont {X.}~\bibnamefont {Li}},\ and\ \bibinfo {author} {\bibfnamefont {X.}~\bibnamefont {Sun}},\ }\bibfield  {title} {\bibinfo {title} {Pairing in the {Hubbard} model on the honeycomb lattice with hopping up to the third-nearest-neighbor},\ }\href {https://doi.org/10.1016/j.physleta.2022.128175} {\bibfield  {journal} {\bibinfo  {journal} {Phys. Lett. A}\ }\textbf {\bibinfo {volume} {442}},\ \bibinfo {pages} {128175} (\bibinfo {year} {2022})}\BibitemShut {NoStop}%
\bibitem [{\citenamefont {Li}\ \emph {et~al.}(2022)\citenamefont {Li}, \citenamefont {Liu}, \citenamefont {Yu}, \citenamefont {Gong}, \citenamefont {Yu},\ and\ \citenamefont {Zhou}}]{liMixtureNearestNextnearestneighbor2022}%
  \BibitemOpen
  \bibfield  {author} {\bibinfo {author} {\bibfnamefont {X.-D.}\ \bibnamefont {Li}}, \bibinfo {author} {\bibfnamefont {H.-R.}\ \bibnamefont {Liu}}, \bibinfo {author} {\bibfnamefont {Z.-D.}\ \bibnamefont {Yu}}, \bibinfo {author} {\bibfnamefont {C.-D.}\ \bibnamefont {Gong}}, \bibinfo {author} {\bibfnamefont {S.-L.}\ \bibnamefont {Yu}},\ and\ \bibinfo {author} {\bibfnamefont {Y.}~\bibnamefont {Zhou}},\ }\bibfield  {title} {\bibinfo {title} {Mixture of the nearest- and next-nearest-neighbor $\textit{d}$ + $\textit{id}$-wave pairings on the honeycomb lattice},\ }\href {https://doi.org/10.1088/1367-2630/ac974a} {\bibfield  {journal} {\bibinfo  {journal} {New J. Phys.}\ }\textbf {\bibinfo {volume} {24}},\ \bibinfo {pages} {103035} (\bibinfo {year} {2022})}\BibitemShut {NoStop}%
\bibitem [{\citenamefont {Shen}\ and\ \citenamefont {Qin}(2024)}]{shenTransitionHalffilledStripe2024}%
  \BibitemOpen
  \bibfield  {author} {\bibinfo {author} {\bibfnamefont {Y.}~\bibnamefont {Shen}}\ and\ \bibinfo {author} {\bibfnamefont {M.}~\bibnamefont {Qin}},\ }\bibfield  {title} {\bibinfo {title} {Transition from half-filled stripe to {N}éel antiferromagnetism in the $t^{\prime}$-{H}ubbard model on the honeycomb lattice},\ }\href {https://doi.org/10.1103/PhysRevB.109.024505} {\bibfield  {journal} {\bibinfo  {journal} {Phys. Rev. B}\ }\textbf {\bibinfo {volume} {109}},\ \bibinfo {pages} {24505} (\bibinfo {year} {2024})}\BibitemShut {NoStop}%
\bibitem [{\citenamefont {Cohen-Stead}\ \emph {et~al.}(2024{\natexlab{a}})\citenamefont {Cohen-Stead}, \citenamefont {Costa}, \citenamefont {Neuhaus}, \citenamefont {Ly}, \citenamefont {Zhang}, \citenamefont {Scalettar}, \citenamefont {Barros},\ and\ \citenamefont {Johnston}}]{10.21468/SciPostPhysCodeb.29}%
  \BibitemOpen
  \bibfield  {author} {\bibinfo {author} {\bibfnamefont {B.}~\bibnamefont {Cohen-Stead}}, \bibinfo {author} {\bibfnamefont {S.~M.}\ \bibnamefont {Costa}}, \bibinfo {author} {\bibfnamefont {J.}~\bibnamefont {Neuhaus}}, \bibinfo {author} {\bibfnamefont {A.~T.}\ \bibnamefont {Ly}}, \bibinfo {author} {\bibfnamefont {Y.}~\bibnamefont {Zhang}}, \bibinfo {author} {\bibfnamefont {R.}~\bibnamefont {Scalettar}}, \bibinfo {author} {\bibfnamefont {K.}~\bibnamefont {Barros}},\ and\ \bibinfo {author} {\bibfnamefont {S.}~\bibnamefont {Johnston}},\ }\bibfield  {title} {\bibinfo {title} {{SmoQyDQMC.jl: A flexible implementation of determinant quantum Monte Carlo for Hubbard and electron-phonon interactions}},\ }\href {https://doi.org/10.21468/SciPostPhysCodeb.29} {\bibfield  {journal} {\bibinfo  {journal} {SciPost Phys. Codebases}\ ,\ \bibinfo {pages} {29}} (\bibinfo {year} {2024}{\natexlab{a}})}\BibitemShut {NoStop}%
\bibitem [{\citenamefont {Cohen-Stead}\ \emph {et~al.}(2024{\natexlab{b}})\citenamefont {Cohen-Stead}, \citenamefont {Costa}, \citenamefont {Neuhaus}, \citenamefont {Ly}, \citenamefont {Zhang}, \citenamefont {Scalettar}, \citenamefont {Barros},\ and\ \citenamefont {Johnston}}]{10.21468/SciPostPhysCodeb.29-r0.3}%
  \BibitemOpen
  \bibfield  {author} {\bibinfo {author} {\bibfnamefont {B.}~\bibnamefont {Cohen-Stead}}, \bibinfo {author} {\bibfnamefont {S.~M.}\ \bibnamefont {Costa}}, \bibinfo {author} {\bibfnamefont {J.}~\bibnamefont {Neuhaus}}, \bibinfo {author} {\bibfnamefont {A.~T.}\ \bibnamefont {Ly}}, \bibinfo {author} {\bibfnamefont {Y.}~\bibnamefont {Zhang}}, \bibinfo {author} {\bibfnamefont {R.}~\bibnamefont {Scalettar}}, \bibinfo {author} {\bibfnamefont {K.}~\bibnamefont {Barros}},\ and\ \bibinfo {author} {\bibfnamefont {S.}~\bibnamefont {Johnston}},\ }\bibfield  {title} {\bibinfo {title} {{Codebase release r0.3 for SmoQyDQMC.jl}},\ }\bibfield  {journal} {\bibinfo  {journal} {SciPost Phys. Codebases}\ }\href {https://doi.org/SciPostPhysCodeb 29-r0.3} {SciPostPhysCodeb 29-r0.3} (\bibinfo {year} {2024}{\natexlab{b}})\BibitemShut {NoStop}%
\bibitem [{\citenamefont {Neuhaus}\ \emph {et~al.}(2024{\natexlab{a}})\citenamefont {Neuhaus}, \citenamefont {Nichols}, \citenamefont {Banerjee}, \citenamefont {Cohen-Stead}, \citenamefont {Maier}, \citenamefont {Maestro},\ and\ \citenamefont {Johnston}}]{10.21468/SciPostPhysCodeb.39}%
  \BibitemOpen
  \bibfield  {author} {\bibinfo {author} {\bibfnamefont {J.}~\bibnamefont {Neuhaus}}, \bibinfo {author} {\bibfnamefont {N.~S.}\ \bibnamefont {Nichols}}, \bibinfo {author} {\bibfnamefont {D.}~\bibnamefont {Banerjee}}, \bibinfo {author} {\bibfnamefont {B.}~\bibnamefont {Cohen-Stead}}, \bibinfo {author} {\bibfnamefont {T.~A.}\ \bibnamefont {Maier}}, \bibinfo {author} {\bibfnamefont {A.~D.}\ \bibnamefont {Maestro}},\ and\ \bibinfo {author} {\bibfnamefont {S.}~\bibnamefont {Johnston}},\ }\bibfield  {title} {\bibinfo {title} {{SmoQyDEAC.jl: A differential evolution package for the analytic continuation of imaginary time correlation functions}},\ }\bibfield  {journal} {\bibinfo  {journal} {SciPost Phys. Codebases}\ }\href {https://doi.org/SciPostPhysCodeb 39} {SciPostPhysCodeb 39} (\bibinfo {year} {2024}{\natexlab{a}})\BibitemShut {NoStop}%
\bibitem [{\citenamefont {Neuhaus}\ \emph {et~al.}(2024{\natexlab{b}})\citenamefont {Neuhaus}, \citenamefont {Nichols}, \citenamefont {Banerjee}, \citenamefont {Cohen-Stead}, \citenamefont {Maier}, \citenamefont {Maestro},\ and\ \citenamefont {Johnston}}]{10.21468/SciPostPhysCodeb.39-r1.1}%
  \BibitemOpen
  \bibfield  {author} {\bibinfo {author} {\bibfnamefont {J.}~\bibnamefont {Neuhaus}}, \bibinfo {author} {\bibfnamefont {N.~S.}\ \bibnamefont {Nichols}}, \bibinfo {author} {\bibfnamefont {D.}~\bibnamefont {Banerjee}}, \bibinfo {author} {\bibfnamefont {B.}~\bibnamefont {Cohen-Stead}}, \bibinfo {author} {\bibfnamefont {T.~A.}\ \bibnamefont {Maier}}, \bibinfo {author} {\bibfnamefont {A.~D.}\ \bibnamefont {Maestro}},\ and\ \bibinfo {author} {\bibfnamefont {S.}~\bibnamefont {Johnston}},\ }\bibfield  {title} {\bibinfo {title} {{Codebase release r1.1 for SmoQyDEAC.jl}},\ }\bibfield  {journal} {\bibinfo  {journal} {SciPost Phys. Codebases}\ }\href {https://doi.org/SciPostPhysCodeb 39-r1.1} {SciPostPhysCodeb 39-r1.1} (\bibinfo {year} {2024}{\natexlab{b}})\BibitemShut {NoStop}%
\bibitem [{\citenamefont {Johnston}\ \emph {et~al.}(2013)\citenamefont {Johnston}, \citenamefont {Nowadnick}, \citenamefont {Kung}, \citenamefont {Moritz}, \citenamefont {Scalettar},\ and\ \citenamefont {Devereaux}}]{johnstonDeterminantQuantumMonte2013}%
  \BibitemOpen
  \bibfield  {author} {\bibinfo {author} {\bibfnamefont {S.}~\bibnamefont {Johnston}}, \bibinfo {author} {\bibfnamefont {E.~A.}\ \bibnamefont {Nowadnick}}, \bibinfo {author} {\bibfnamefont {Y.~F.}\ \bibnamefont {Kung}}, \bibinfo {author} {\bibfnamefont {B.}~\bibnamefont {Moritz}}, \bibinfo {author} {\bibfnamefont {R.~T.}\ \bibnamefont {Scalettar}},\ and\ \bibinfo {author} {\bibfnamefont {T.~P.}\ \bibnamefont {Devereaux}},\ }\bibfield  {title} {\bibinfo {title} {Determinant quantum {{Monte Carlo}} study of the two-dimensional single-band {{Hubbard-Holstein}} model},\ }\href {https://doi.org/10.1103/PhysRevB.87.235133} {\bibfield  {journal} {\bibinfo  {journal} {Phys. Rev. B}\ }\textbf {\bibinfo {volume} {87}},\ \bibinfo {pages} {235133} (\bibinfo {year} {2013})},\ \Eprint {https://arxiv.org/abs/1306.2968} {1306.2968} \BibitemShut {NoStop}%
\bibitem [{\citenamefont {Li}\ \emph {et~al.}(2002)\citenamefont {Li}, \citenamefont {Rignanese}, \citenamefont {Chang}, \citenamefont {Blase},\ and\ \citenamefont {Louie}}]{liGWStudyMetalinsulator2002}%
  \BibitemOpen
  \bibfield  {author} {\bibinfo {author} {\bibfnamefont {J.-L.}\ \bibnamefont {Li}}, \bibinfo {author} {\bibfnamefont {G.-M.}\ \bibnamefont {Rignanese}}, \bibinfo {author} {\bibfnamefont {E.~K.}\ \bibnamefont {Chang}}, \bibinfo {author} {\bibfnamefont {X.}~\bibnamefont {Blase}},\ and\ \bibinfo {author} {\bibfnamefont {S.~G.}\ \bibnamefont {Louie}},\ }\bibfield  {title} {\bibinfo {title} {{{GW}} study of the metal-insulator transition of bcc hydrogen},\ }\href {https://doi.org/10.1103/PhysRevB.66.035102} {\bibfield  {journal} {\bibinfo  {journal} {Phys. Rev. B}\ }\textbf {\bibinfo {volume} {66}},\ \bibinfo {pages} {035102} (\bibinfo {year} {2002})}\BibitemShut {NoStop}%
\bibitem [{\citenamefont {Berger}\ \emph {et~al.}(1995)\citenamefont {Berger}, \citenamefont {Valášek},\ and\ \citenamefont {Von Der~Linden}}]{bergerTwodimensionalHubbardHolsteinModel1995}%
  \BibitemOpen
  \bibfield  {author} {\bibinfo {author} {\bibfnamefont {E.}~\bibnamefont {Berger}}, \bibinfo {author} {\bibfnamefont {P.}~\bibnamefont {Valášek}},\ and\ \bibinfo {author} {\bibfnamefont {W.}~\bibnamefont {Von Der~Linden}},\ }\bibfield  {title} {\bibinfo {title} {Two-dimensional {{Hubbard-Holstein}} model},\ }\href {https://doi.org/10.1103/PhysRevB.52.4806} {\bibfield  {journal} {\bibinfo  {journal} {Phys. Rev. B}\ }\textbf {\bibinfo {volume} {52}},\ \bibinfo {pages} {4806} (\bibinfo {year} {1995})}\BibitemShut {NoStop}%
\bibitem [{\citenamefont {Liu}\ \emph {et~al.}()\citenamefont {Liu}, \citenamefont {Zhang},\ and\ \citenamefont {Ma}}]{liu_2026_18616136}%
  \BibitemOpen
  \bibfield  {author} {\bibinfo {author} {\bibfnamefont {H.}~\bibnamefont {Liu}}, \bibinfo {author} {\bibfnamefont {L.}~\bibnamefont {Zhang}},\ and\ \bibinfo {author} {\bibfnamefont {T.}~\bibnamefont {Ma}},\ }\bibfield  {title} {\bibinfo {title} {Data associated with tuning superconductivity and charge-density-wave order by next-nearest-neighbor hopping integral in honeycomb holstein model {[Dataset]}, {Zenodo} (2026)},\ }\href {https://doi.org/https://doi.org/10.5281/zenodo.18616136} {https://doi.org/10.5281/zenodo.18616136}\BibitemShut {NoStop}%
\end{thebibliography}%

\end{document}